\documentclass[11pt]{amsart}
\usepackage[utf8]{inputenc}
\usepackage[titletoc]{appendix}

\usepackage[backend=biber, style=numeric, date=year, url=false, doi=false, isbn=false, eprint=true, firstinits=true, maxcitenames=5, maxbibnames=5]{biblatex}
\AtEveryBibitem{%
  \ifentrytype{online}{%
  }{%
    \clearfield{eprint}%
    \clearfield{urldate}%
  }%
}
\usepackage{blindtext}
\usepackage{amssymb}
\usepackage{amsfonts}
\usepackage{amsthm}
\usepackage{tikz}
\usepackage{caption}
\usepackage{graphics}
\usepackage{xcolor}
\usepackage{shadethm}
\usepackage{subfigure}
\usepackage{epigraph}
\usepackage[includeheadfoot,margin=1.2in]{geometry}
\usepackage{placeins}
\usetikzlibrary{snakes}
\usepackage[new]{old-arrows}
\usepackage{newtxtext,newtxmath}

\usepackage{enumerate}
\usepackage{verbatim}
\usetikzlibrary{trees}
\usetikzlibrary{decorations.markings}
\usetikzlibrary{arrows}
\usetikzlibrary{cd}
\usepackage{mathrsfs}
\usepackage{xparse}
\usetikzlibrary{positioning,arrows,patterns}
\usetikzlibrary{decorations.markings}
\usetikzlibrary{calc}
\tikzset{
  photon/.style={decorate, decoration={snake}, draw=black},
  fermion/.style={draw=black, postaction={decorate},decoration={markings,mark=at position .55 with {\arrow{>}}}},
  fermion2/.style={dashed, dash phase=0.1pt, draw=black, postaction={decorate},decoration={markings,mark=at position .55 with {\arrow{>}}}},
  vertex/.style={draw,shape=circle,fill=black,minimum size=5pt,inner sep=0pt},
particle/.style={thick,draw=black},
particle2/.style={thick,draw=blue},
avector/.style={thick,draw=black, postaction={decorate},
    decoration={markings,mark=at position 1 with {\arrow[black]{triangle 45}}}},
gluon/.style={decorate, draw=black,
    decoration={coil,aspect=0}}
 }
\NewDocumentCommand\semiloop{O{black}mmmO{}O{above}}
{%
\draw[#1] let \p1 = ($(#3)-(#2)$) in (#3) arc (#4:({#4+180}):({0.5*veclen(\x1,\y1)})node[midway, #6] {#5};)
}

\usepackage[mathcal]{euscript}

\usepackage{url}
\usepackage{hyperref}
\hypersetup{colorlinks=true, citecolor=purple, linktocpage, linkcolor=blue, urlcolor=cyan} 

\theoremstyle{plain}
\newtheorem{thm}{Theorem}[section]
\newtheorem{lem}[thm]{Lemma}
\newtheorem{prop}[thm]{Proposition}

\theoremstyle{definition}

\newtheorem*{thm*}{Theorem}
\newtheorem*{lem*}{Lemma}
\newtheorem*{prop*}{Proposition}
\newtheorem*{cor*}{Corollary}
\newtheorem*{exe*}{Exercise}
\newtheorem*{defn*}{Definition}
\newtheorem{rem}[thm]{Remark}

\newtheorem{ex}[thm]{Example}

\theoremstyle{remark}

\newcommand{\R}{\mathbb{R}}
\newcommand{\Z}{\mathbb{Z}}

\newcommand{\E}{\mathbb{E}}

\newcommand{\calU}{\mathcal{U}}
\newcommand{\calD}{\mathcal{D}}

\newcommand{\dd}{{\mathrm{d}}}

\newcommand{\calR}{\mathcal{R}}

\newcommand{\id}{\mathrm{id}}

\DeclareMathOperator{\tr}{Tr}

\DeclareMathOperator{\Div}{\textnormal{div}}

\DeclareMathOperator{\gh}{gh}

\newcommand{\Der}{{\mathrm{Der}}}

\DeclareMathOperator{\End}{End}
\DeclareMathOperator{\Hom}{Hom}

\newcommand{\T}{\textsf{T}}

\newcommand{\de}{\partial}

\newcommand{\calB}{\mathcal{B}}
\newcommand{\calH}{\mathcal{H}}

\newcommand{\calC}{\mathcal{C}}

\newcommand{\calG}{\mathcal{G}}

\newcommand{\calO}{\mathcal{O}}
\newcommand{\calL}{\mathcal{L}}
\newcommand{\calM}{\mathcal{M}}

\newcommand{\calW}{\mathcal{W}}
\newcommand{\calT}{\mathcal{T}}

\newcommand{\calF}{\mathcal{F}}

\def\gpd{\,\lower1pt\hbox{$\longrightarrow$}\hskip-.24in\raise2pt
               \hbox{$\longrightarrow$}\,}

\let\Tilde=\widetilde

\let\Hat=\widehat

\newcommand{\hateta}{\widehat{\boldsymbol\eta}}
\newcommand{\hatX}{\widehat{\mathsf{X}}}

\DeclareMathOperator{\Map}{Map}

\newcommand{\I}{\mathrm{i}}

\newcommand{\calV}{\mathcal{V}}

\newcommand{\Sym}{\textnormal{Sym}}
\newcommand{\Tr}{\textnormal{Tr}}
\makeatletter
\newcommand{\upint}{\DOTSI\upintop\ilimits@}
\newcommand{\upoint}{\DOTSI\upointop\ilimits@}
\makeatother

\usepackage{todonotes}

\makeatletter
\providecommand\@dotsep{5}
\renewcommand{\listoftodos}[1][\@todonotes@todolistname]{%
  \@starttoc{tdo}{#1}}
\makeatother

\makeatletter

\pgfdeclareshape{genus}{
 \anchor{center}{\pgfpointorigin}
\backgroundpath{
    \begingroup
    \tikz@mode
    \iftikz@mode@fill

         \pgfpathmoveto{\pgfqpoint{-0.78cm}{-.17cm}}
     \pgfpathcurveto %
            {\pgfpoint{-0.35cm}{-.44cm}}
        {\pgfpoint{0.35cm}{-.44cm}}
        {\pgfpoint{.78cm}{-0.17cm}} 
     \pgfpathmoveto{\pgfqpoint{-0.78cm}{-0.17cm}}
     \pgfpathcurveto %
            {\pgfpoint{-0.25cm}{.25cm}}
        {\pgfpoint{.25cm}{.25cm}}
        {\pgfpoint{0.78cm}{-0.17cm}}
        \pgfusepath{fill}
        \fi

        \iftikz@mode@draw
     \pgfpathmoveto{\pgfqpoint{-1cm}{0cm}}
     \pgfpathcurveto %
            {\pgfpoint{-0.5cm}{-.5cm}}
        {\pgfpoint{0.5cm}{-.5cm}}
        {\pgfpoint{1cm}{0cm}}

     \pgfpathmoveto{\pgfqpoint{-0.75cm}{-0.15cm}}
     \pgfpathcurveto %
            {\pgfpoint{-0.25cm}{.25cm}}
        {\pgfpoint{.25cm}{.25cm}}
        {\pgfpoint{0.75cm}{-0.15cm}}
              \pgfusepath{stroke}
        \fi
        \endgroup
    }
    }

    \makeatother

\title[On Globalized Traces for the Poisson Sigma Model]{On Globalized Traces for the Poisson Sigma Model}
\author[N. Moshayedi]{Nima Moshayedi}
\address{Institut f\"ur Mathematik\\ Universit\"at Z\"urich\\ 
Winterthurerstrasse 190
CH-8057 Z\"urich}
\email[N.~Moshayedi]{nima.moshayedi@math.uzh.ch}
\dedicatory{Dedicated to Giovanni Felder on the occasion of his 60th birthday}

\begin{document}

\maketitle


\tikzset{residual/.style={draw, shape=circle, black,inner sep=1pt}}

\begin{abstract}
A globalized version of a trace formula for the Poisson Sigma Model on the disk is presented by using its formal global picture in the setting of the Batalin--Vilkovisky formalism. This global construction includes the concept of zero modes. Moreover, for the symplectic case of the Poisson Sigma Model with cotangent target, the globalized trace reduces to a symplectic construction which was presented by Grady, Li and Li in \cite{GLL} for 1-dimensional Chern--Simons theory (topological quantum mechanics). In addition, the connection between this formula and the Nest--Tsygan theorem and the Tamarkin--Tsygan theorem is explained.


\end{abstract}

\tableofcontents

\section{Introduction}
In \cite{K} Kontsevich showed that the differential graded Lie algebra (DGLA) of multidifferential operators on a manifold $M$ is $L_\infty$-quasi-isomorphic to the DGLA of multivector fields on $M$. This is known as the formality theorem.
The construction of Kontsevich's star product in deformation quantization is given by the special case of the formality theorem for bivector fields and bidifferential operators. In \cite{CF1} it was shown that this star product can be written as a perturbative expansion of a path integral given by the Poisson Sigma Model \cite{SS1,I}. In \cite{Tsygan1999} Tsygan formulated a formality conjecture for cyclic chains (which was motivated as a chain version of the Connes--Flato--Sternheimer cyclic cohomology construction \cite{ConnesFlatoSternheimer1992}), which was partially proven by Shoikhet \cite{Shoikhet2003}, Dolgushev \cite{Dolgushev2006} and Willwacher \cite{Willwacher2011}. In \cite{WillwacherCalaque2012} Willwacher and Calaque have proven the cyclic formaility conjecture of Kontsevich, which was the formulation for cyclic cochains. 

A global geometrical picture of the star product coming from the Weyl quantization approach for symplectic manifolds, i.e. for a constant Poisson structure, was given by Fedosov in \cite{Fedosov1994}. 
There one chooses a(n) (always existing) symplectic connection and its corresponding exponential map. This construction can be generalized to the local picture of Kontsevich's star product to produce a global version on any Poisson manifold \cite{CFT,CattaneoFelderTomassini2002}, where one uses notions of formal geometry \cite{B,GK}. The symplectic connection (lifted to the Weyl bundle) can be replaced by the (deformed) Grothendieck connection which is constructed by using any (formal) exponential map (see also \cite{CF3}).
A globalized picture in the field theoretic approach using the Poisson Sigma Model in the Batalin--Vilkovisky (BV) formalism \cite{BV1,BV2,BV3} was given in \cite{BCM} for closed worldsheet manifolds and in \cite{CMW3} for manifolds with boundary using the BV-BFV formalism \cite{CMR1,CMR2,CMW4,CattMosh1}. Here BFV stands for Batalin--Fradkin--Vilkovisky, which is the Hamiltonian approach of the BV formalism developed in \cite{BF1,BF2}.

An important object to study for closed\footnote{Let $A=C^\infty(M)$ for some Poisson manifold $M$. For a deformed algebra $(A[\![\hbar]\!],\star)$, closedness means that the integration of elements of the deformed algebra is a trace with respect to $\star$. In particular, a star product is closed if and only if $\int_M (f\star g)\Omega=\int_M(g\star f)\Omega$ for any $f,g\in A$, where $\Omega$ is a volume form on $M$.} star products \cite{ConnesFlatoSternheimer1992} are trace maps. In \cite{NestTsygan1995} Nest and Tsygan showed an algebraic version of the Atiyah--Singer index theorem, where they made the link to a trace map with respect to the underlying star product and computed the index as the trace of the constant function 1 (see also \cite{Fedosov1996} for Fedosov's construction). This construction is given for symplectic manifolds together with the globalization construction of the Moyal product. For a general Poisson manifold with Kontsevich's star product, Cattaneo and Felder constructed a trace map in terms of local field theoretic constructions using the Poisson Sigma Model on the disk for negative cyclic chains \cite{CattaneoFelder2010} in the presence of residual fields (a.k.a ``slow'' fields,``low energy'' fields). 

We will extend this construction to a global one by using a formal global version of the Poisson Sigma Model. This construction in fact combines Fedosov's globalization construction with field theoretic concepts on Poisson manifolds and the BV formulation. We also give the connection of the obtained globalized trace to the Tamarkin--Tsygan theorem, which can be seen as a cyclic equivariant extension of the Nest--Tsygan theorem for Poisson manifolds using formally extended Poisson structures. The connection can be understood by field theoretic concepts by looking at the Feynman graph expansion for the obtained trace formula, which geometrically gives rise to a deformed version of the Grothendieck connection and its curvature. 

In \cite{GLL} a global equivariant trace formula for symplectic manifolds was constructed by using a Fedosov connection and solutions to the Fedosov equation. The field theoretic construction was given by the effective theory of topological quantum mechanics on the circle $S^1$. We show that our trace formula reduces to this trace formula if we consider the Poisson Sigma Model with cotangent target. To show this, we use the fact (Proposition \ref{prop:linear_R}) that the vertices of our graphs in the expansion which arise from the Grothendieck connection are linear in the fiber coordinates if the underlying manifold is a cotangent bundle.

\subsection*{Acknowledgements}
I would like to thank Alberto Cattaneo for introducing me to this problem, for several discussions and comments. I would also like to thank Giovanni Felder and Thomas Willwacher for short discussions at Monte Verit\`a and for the organization of a great conference. Moreover, I would like to thank Konstantin Wernli and Nicola Capacci for discussions at different stages of this paper.

This research was (partly) supported by the NCCR SwissMAP, funded by the Swiss National Science Foundation. I
acknowledge partial support of SNF grant No. 200020\_172498/1.

\section{Cyclic formality}

\subsection{The Kontsevich Formality}
Let $(\calT_{poly}^\bullet(\R^d),[\enspace,\enspace]_{SN},\dd=0)$ be the DGLA of multivector fields on $\R^d$ endowed with the Schouten--Nijenhuis bracket and the zero differential and let $(\calD_{poly}^\bullet(\R^d),[\enspace,\enspace]_G,b)$ be the DGLA of multidifferential operators on $\R^d$ endowed with the Gerstenhaber bracket and the Hochschild differential.
In \cite{K} Kontsevich proved the celebrated formality theorem, which states that these two complexes are quasi-isomorphic as $L_\infty$-algebras. 
\begin{thm}[Kontsevich \cite{K}]
\label{thm:formality}
There exists an $L_\infty$-quasi-isomorphism
\begin{equation}
\calU\colon(\calT_{poly}^\bullet(\R^d),[\enspace,\enspace]_{SN},\dd=0)\longrightarrow(\calD_{poly}^\bullet(\R^d),[\enspace,\enspace]_G,b).
\end{equation}
\end{thm}
For the case of degree two, Theorem \ref{thm:formality} implies a star product on $\R^d$ endowed with any Poisson structure.
Moreover, Theorem \ref{thm:formality} can be extended to a global version, where $\R^d$ can be replaced by any finite-dimensional manifold $M$ as we will also describe in Section \ref{subsec:globalization}. Let us briefly recall the main objects to understand the \emph{\textbf{formality theorem}}.

\subsubsection{The Hochschild complex and the Gerstenhaber bracket}
Let $A$ be a unital algebra with unit $1$. One can consider the graded algebra $C_\bullet(A):=A\otimes \bar A^{\otimes \bullet}$, where $\bar A:=A/\R 1$. This space is endowed with a map 
\begin{equation}
b([a_0\otimes\dotsm\otimes a_m])=\sum_{i=1}^{m-1}(-1)^{i}[a_0\otimes\dotsm\otimes a_ia_{i+1}\otimes\dotsm\otimes a_m]+(-1)^m[a_ma_0\otimes\dotsm \otimes a_{m-1}].
\end{equation}
One can check that $b\colon C_\bullet(A)\longrightarrow C_{\bullet-1}(A)$ is a differential, called the \emph{\textbf{Hochschild differential}} and the tuple $(C_\bullet(A),b)$ is called the \emph{\textbf{Hochschild chain complex}} of $A$. Here we denote by $[a_0\otimes\dotsm \otimes a_m]$ the class of $a_0\otimes\dotsm \otimes a_m$ in $C_\bullet(A)$. Moreover, we define $C_m(A)=0$ for all $m<0$. 
The DGLA of multidifferential operators $\calD_{poly}^\bullet(M)$, for a manifold $M$, can thus be seen as the subcomplex of the shifted complex $C^\bullet(A):=\Hom(A^{\otimes\bullet+1},A)$, where $A=C^\infty(M)$, consisting of multilinear maps which are differential operators in each argument.
The \emph{\textbf{Gerstenhaber bracket}} of two multidifferential operators $D,D'$ is given by
\begin{equation}
[D,D']_G:=D\bullet_GD'-(-1)^{\vert D\vert\cdot\vert D'\vert}D'\bullet_GD,
\end{equation}
where $\vert D\vert$ denotes the degree of the multidifferential operator $D$ and the \emph{\textbf{Gerstenhaber product}} $\bullet_G$ is given by
\begin{equation}
D\bullet_G D':=\sum_{k=0}^n(-1)^{\vert D'\vert\cdot(\vert D\vert-k)}D\circ (\id^{\otimes k}\otimes D'\otimes \id^{\otimes \vert D\vert-k}).
\end{equation}
The differential on $C^\bullet(A)$ is given in terms of the Gerstenhaber bracket by $[\mu,\enspace]_G$ for $\mu\in \Hom(A\otimes A,A)$ being the multiplication map of $A$. In fact, in \cite{Ger1963} it was shown that the Hochschild cohomology $HH^\bullet(A)$ together with $\bullet_G$ and $[\enspace,\enspace]_G$ is a Gerstenhaber algebra.

\subsubsection{Multivector fields and the Schouten--Nijenhuis bracket}
The space of multivector fields on a manifold $M$ is given by $\Gamma(\bigwedge^\bullet TM)$. 
We define $\calT^{\bullet}_{poly}(M):=\bigoplus_{j\geq -1}\Gamma(\bigwedge^{j+1}TM)$, with the convention that $\calT^{-1}_{poly}(M)=C^\infty(M)$, $\calT^{0}_{poly}(M)=\Gamma(TM)$, $\calT^{1}_{poly}(M)=\Gamma(\bigwedge^2TM)$, etc.
The \emph{\textbf{Schouten--Nijenhuis bracket}} $[\enspace,\enspace]_{SN}$ is given by the usual Lie bracket extended to multivector fields by the Leibniz rule, i.e. for multivector fields $\alpha,\beta,\gamma$ we have
\begin{equation}
[\alpha\land\beta,\gamma]_{SN}=\alpha\land[\beta,\gamma]_{SN}+(-1)^{\vert\gamma\vert\cdot(\vert\beta\vert+1)}[\alpha,\gamma]_{SN}\land\beta.
\end{equation}

\subsubsection{The Hochschild--Kostant--Rosenberg map}
Consider vector fields $\xi_1,\ldots,\xi_n\in \calT_{poly}^0(M)$ and $f_1,\ldots,f_n\in A$. One can construct a map, which for $n\geq 1$ is given by
\begin{align}
\begin{split}
\calT_{poly}^{n-1}(M)&\longrightarrow \calD_{poly}^{n-1}(M)\\
\xi_1\land\dots\land\xi_n&\longmapsto \left(f_1\otimes\dotsm \otimes f_n\longmapsto \frac{1}{n!}\sum_{\sigma\in S_n}\textnormal{sign}(\sigma)\xi_{\sigma(1)}(f_1)\dotsm \xi_{\sigma(n)}(f_n)\right),
\end{split}
\end{align}
and for $n=0$ it is given by the identity on $C^\infty(M)$.
Here $S_n$ denotes the symmetric group of order $n$.
This map is called \emph{\textbf{Hochschild--Kostant--Rosenberg (HKR) map}}. One can check that it is indeed a chain map and a quasi-isomorphism of complexes, but does not respect the Lie bracket on the level of complexes. In fact Kontsevich's $L_\infty$-quasi-isomorphism $\calU$ gives a solution to this problem as a certain extension of the HKR map. In particular, the first Taylor component $\calU_1$ of $\calU$ is precisely the HKR map.

\subsection{The Kontsevich--Tsygan Formality}
\label{subsec:Kontsevich_Tsygan_formality}
One can generalize the formality construction to a cyclic version by considering \emph{\textbf{cyclic chains}}. There is another differential, called the \emph{\textbf{Connes differential}} \cite{Connes1985,ConnesFlatoSternheimer1992}, of degree $+1$ on the Hochschild complex given by
\begin{equation}
B([a_0\otimes\dotsm\otimes a_m]):=\sum_{i=0}^{m}(-1)^{im}[1\otimes a_i\otimes\dotsm \otimes a_m\otimes a_0\otimes\dotsm \otimes a_{i-1}].
\end{equation}
Note that there is an HKR chain map 
\begin{align}
\begin{split}
    (C_\bullet(A),b)&\longrightarrow (\Omega^\bullet(M,\R),\dd=0)\\
    [a_0\otimes\dotsm\otimes a_m]&\longmapsto \frac{1}{m!}a_0\dd a_1\land\dotsm \land\dd a_m.
\end{split}
\end{align}
This map is also called the \emph{\textbf{Connes map}} \cite{ConnesFlatoSternheimer1992}, which identifies cyclic and de Rham cohomology.
Following Getzler \cite{Getzler1993}, the \emph{\textbf{negative cyclic chain complex}} is then given by 
\begin{equation}
CC_{-\bullet}^-(A):=C_{-\bullet}(A)[u]
\end{equation}
endowed with the differential $b+uB$. Here $u$ denotes some formal variable of degree 2. Similarly to the negative cyclic chain complex, one can define the \emph{\textbf{periodic cyclic chain complex}} by allowing negative powers of the formal parameter $u$, hence we have the formal Laurent polynomials $PC_{-\bullet}(A):=C_{-\bullet}(A)[u,u^{-1}]$. 
We can extend the HKR map by $\R[u]$-linearity and obtain a quasi-isomorphism
\begin{equation}
(CC^-_{-\bullet}(A),b+uB)\longrightarrow (\Omega^{-\bullet}(M,\R)[u],u\dd).
\end{equation}
Consider a module $W$ over the graded algebra $\R[u]$ of finite projective dimension and define $CC_{-\bullet}^W(A):=C_{-\bullet}(A)[u]\otimes_{\R[u]}W$.
The formality for cyclic chains is given by the following theorem.
\begin{thm}[Kontsevich--Tsygan \cite{Tsygan1999}]
\label{thm:cyclic_chain_formality}
There exists an $L_\infty$-quasi-isomorphism 
\begin{equation}
\label{eq:cyclic_chain_formality}
\calU^{cyc}\colon (CC_{-\bullet}^W(A),b+uB)\longrightarrow (\Omega^{-\bullet}(M,\R)[u]\otimes_{\R[u]}W,u\dd).
\end{equation}
\end{thm}
This was proven by Shoikhet, Willwacher and globally extended by Dolgushev using Fedosov resolution. Using Shoikhet's $L_\infty$-quasi-isomorphism $\calU^{Sh}$, one can obtain Theorem \ref{thm:cyclic_chain_formality} as a corollary by obtaining $\calU^{Sh}\circ b=\dd \circ \calU^{Sh}$ \cite{Willwacher2011}.

\begin{rem}
This construction leads to a field theoretic construction using the Poisson Sigma Model on the disk as we will see in Section \ref{sec:traces_and_algebraic_index}. One can construct a trace map which uses an $\R[u]$-linear morphism of $L_\infty$-modules over some suitable algebra. 
\end{rem}

\section{Fedosov's approach to deformation quantization}
In this section we want to recall the most important notions and constructions of \cite{Fedosov1994}. 
\subsection{Weyl algebra and Moyal product}
Let $(M,\omega)$ be a symplectic manifold and let $\{x_i\}$ be local coordinates on $M$ and $\{y^i\}$ coordinates on the corresponding fiber of the tangent bundle, i.e. $(x^{i},y^{i})\in M\times T_{x^i}M$. Consider the \emph{\textbf{Weyl bundle}} $\calW(M):=\Hat{\Sym}(T^*M)[\![ \hbar ]\!]$ associated to $M$, where $\Hat{\Sym}$ denotes the completed symmetric algebra. The Weyl bundle can be regarded as a deformation  of the bundle of formal functions on $T^*M$.
We will write $\calW$ instead of $\calW(M)$ whenever it is clear.
A section $a\in\Gamma(\calW)$ is locally given by\footnote{We will use the Einstein summation convention.} 
\begin{equation}
a(x,y,\hbar)=\sum_{k,\ell}\hbar^k a_{k,i_1,\ldots,i_\ell}(x)y^{i_1}\dotsm y^{i_\ell},
\end{equation}
where $a_{k,i_1,\ldots,i_\ell}\in C^\infty(M)$. In each fiber $\calW_x$ for $x\in M$, one can construct an algebra structure by considering the associative product
\begin{align}
\begin{split}
\star\colon \calW_x\times \calW_x&\longrightarrow \calW_x,\\
(a(x,\hbar),b(x,\hbar))&\longmapsto (a\star b)(x,\hbar):=\exp\left(-\frac{\I\hbar}{2}\omega^{ij}\frac{\de}{\de y^{i}}\frac{\de}{\de z^{j}}\right)a(y,\hbar)b(z,\hbar)\Big|_{z=y}\\
&\qquad = \sum_{k=0}^\infty \left(-\frac{\I\hbar}{2}\right)^k\frac{1}{k!}\omega^{i_1j_1}\dotsm \omega^{i_kj_k}\frac{\de^k a}{\de y^{i_1}\dotsm \de y^{i_k}}\frac{\de^k b}{\de z^{j_1}\dotsm \de z^{j_k}}.
\end{split}
\end{align}
Here we denote by $(\omega^{ij})$ the components of the inverse $\omega^{-1}$ of the symplectic form. For any $x\in M$, the tuple $(\calW_x,\star)$ is called the \emph{\textbf{Weyl algebra}} and $\star$ is called the \emph{\textbf{Moyal product}}. One can check that
\begin{equation}
\lim_{\hbar\to 0}\frac{1}{\hbar}(a\star b-b\star a)=\{a,b\},
\end{equation}
where $\{\enspace,\enspace\}$ is the Poisson bracket coming from the symplectic structure $\omega$, which makes sure that $\star$ is actually a deformation quantization of $T^*_xM$ with constant Poisson structure $\omega_x^{-1}$.
Let $\Omega^\bullet(M,\calW)$ denote the space of global differential forms on $M$ with values in $\calW$. A section $a\in \Gamma(\Omega^\bullet(M,\calW))$ is of the form 
\begin{equation}
a(x,y,\dd x,\hbar)=\sum_{k,p,q}\hbar^k a_{k,i_1,\ldots,i_p,j_1,\ldots,j_q}(x)y^{i_1}\dotsm y^{i_p}\dd x^{j_1}\land\dotsm \land\dd x^{j_q},
\end{equation} 
Moreover, we define the operators $\delta$ and $\delta^*$ according to \cite{Fedosov1994} by 
\begin{equation}
\delta a:=\dd x^k\land \frac{\de a}{\de y^k},\qquad \delta^* a:=y^k\iota_{\frac{\de}{\de x^k}}a.
\end{equation}
where $\iota$ denotes the contraction. Define $\delta^{-1}:=\frac{1}{p+q}\delta^*$ for $p+q>0$ and zero if $p+q=0$. 
\subsection{Symplectic connection and curvature}
Consider now a symplectic connection $\nabla^{TM}$ on the tangent bundle $TM$, i.e. a torsion-free connection such that $\nabla^{TM}\omega=0$. This induces directly a connection $\nabla^\calW$ on $\calW$ which we will just denote by $\nabla$. The curvature of this connection is given by
\begin{equation}
F^\nabla=\frac{1}{2}F^{i}_{jk\ell}\dd x^{k}\land \dd x^\ell.    
\end{equation}
Moreover, consider the tensor
\begin{equation}
\label{eq:Weyl_curvature}
F:=\frac{1}{4}F_{ijk\ell}y^{i}y^j\dd x^k\land \dd x^\ell,\quad F_{ijk\ell}:=\omega_{im} F_{jk\ell}^m.
\end{equation}
In \cite{Fedosov1994} it was shown that the curvature of $\nabla$ can be formulated as 
\begin{equation}
\label{eq:curv_fedosov}
\nabla^2=\frac{1}{\hbar}[F,\enspace]_\star,
\end{equation}
where $[\enspace,\enspace]_\star$ denotes the commutator with respect to the Moyal product $\star$. \subsection{Fedosov's main theorems}
Consider a connection 
\begin{equation}
\label{eq:fed_conn}
\bar\nabla :=\nabla +\frac{1}{2\hbar}[\gamma,\enspace]_\star
\end{equation}
on $\calW$, where $\gamma\in\Omega^1(M,\calW)$. One can check that $\bar\nabla$ is compatible with the Moyal product, i.e.
\begin{equation}
\label{eq:Moyal_prod_comp}
\bar\nabla(a\star b)=\bar\nabla(a)\star b+ a\star \bar\nabla(b).
\end{equation}

\begin{thm}[Fedosov \cite{Fedosov1994}]
Consider a sequence $\{\omega_k\}_{k\geq 1}$ of closed 2-forms on $M$. Then there is a flat connection $\bar\nabla$ (that is $\bar\nabla^2=0$) defined as in \eqref{eq:fed_conn} such that $\gamma=\sum_{i,j}\omega_{ij}y^{i}\dd x^j+r$, where $r\in \Omega^1(M,\calW)$ satisfying $\delta^{-1}r=0$. Moreover, $\gamma$ satisfies 
\begin{equation}
\label{eq:fedosov}
\bar\nabla\gamma=\nabla\gamma+\frac{1}{2\hbar}[\gamma,\gamma]_\star+F=\omega_\hbar,\quad \omega_\hbar:=-\omega+\sum_{k\geq 1}\hbar^k\omega_k.   
\end{equation}
\end{thm}
Consider the \emph{\textbf{symbol map}} 
\begin{align}
\label{eq:symbol_map}
\begin{split}
\sigma\colon \Gamma(\calW)&\longrightarrow C^\infty(M)[\![ \hbar ]\!]\\
a(x,y,\hbar)=\sum_{k,\ell}\hbar^k a_{k,i_1,\ldots,i_\ell}(x)y^{i_1}\dotsm y^{i_\ell}&\longmapsto a(x,0,\hbar)=\sum_{k}\hbar^k a_{k,i_1,\ldots,i_\ell}(x),
\end{split}
\end{align}
which sends all the $y^{i}$s to zero. 
\begin{thm}[Fedosov \cite{Fedosov1994}]
\label{thm:fedosov}
The symbol map induces an isomorphism 
\begin{equation}
\sigma\colon H^0_\nabla(\Gamma(\calW))\xrightarrow{\sim}C^\infty(M)[\![ \hbar ]\!], 
\end{equation}
where $H^0_\nabla(\Gamma(\calW))$ denotes the space of flat sections of the Weyl bundle with respect to $\nabla$. Moreover, since for any flat connection Equation \eqref{eq:Moyal_prod_comp} holds, we can construct a global star product on $C^\infty(M)[\![ \hbar ]\!]$ by the formula 
\begin{equation}
f\star _Mg:=\sigma(\sigma^{-1}(f)\star \sigma^{-1}(g)),
\end{equation}
which defines a deformation quantization on $(M,\omega)$.
\end{thm}
\begin{rem}
Theorem \ref{thm:fedosov} tells us the existence of a global version of the Moyal product for symplectic manifolds. There is a similar approach to globalization for any Poisson manifold, where we start with Kontsevich's star product on the local picture using elements of formal geometry, such as the construction of the Grothendieck connection. A modification (deformed version) of this connection  will replace the symplectic connection in Fedosov's picture. In fact, Fedosov's construction uses the exponential map of a symplectic connection, whereas the more general approach uses the notion of a formal exponential map as we will discuss in the next section.
\end{rem}

\section{Formal geometry and Grothendieck connection}
In this section we want to recall the most important notions of formal geometry as in \cite{B,GK}, the construction of the Grothendieck connection, its deformed version and the relation to Fedosov's quantization approach for the case of a symplectic manifold \cite{CF3,CFT,CattaneoFelderTomassini2002,BCM,CMW3,CMW4}. 

\subsection{Formal exponential maps}
Let $M$ be a smooth manifold. Let $\varphi \colon U \longrightarrow M$ where $U \subset TM$ is an open neighbourhood of the zero section. For $x \in M, y \in T_xM \cap U$ we write $\varphi_x(y):=\varphi(x,y)$. We say that $\varphi$ is a \emph{\textbf{generalized exponential map}} if for all $x \in M$ we have that $\varphi_x(0) = x$, and $\dd\varphi_x\vert_{y=0} = \mathrm{id}_{T_xM}$. In local coordinates we can write 
\begin{equation}
\varphi_x^{i}(y)=x^{i}+y^{i}+\frac{1}{2}\varphi_{x,jk}^{i}y^jy^k+\frac{1}{3!}\varphi^{i}_{x,jk\ell}y^jy^ky^\ell+\dotsm
\end{equation}

where the $x^i$ are coordinates on the base and the $y^i$ are coordinates on the fibers. 
We identify two generalized exponential maps if their jets at $y=0$ agree to all orders. A \emph{\textbf{formal exponential map}} is an equivalence class of generalized exponential maps. It is completely specified by the sequence of functions $\left(\varphi^i_{x,i_1,\ldots,i_k}\right)_{k=0}^{\infty}$. By abuse of notation, we will denote equivalence classes and their representatives by $\varphi$. From a formal exponential map $\varphi$ and a function $f \in C^{\infty}(M)$, we can produce a section $\sigma \in \Gamma(\Hat{\Sym}(T^*M))$ by defining $\sigma_x = \T\varphi_x^*f$, where $\T$ denotes the Taylor expansion in the fiber coordinates around $y=0$ and we use any representative of $\varphi$ to define the pullback. We denote this section by $\T\varphi^*f$; it is independent of the choice of representative, since it only depends on the jets of the representative. 

\begin{ex}
The exponential map of a connection is an example of an exponential map.
\end{ex}

\subsection{The Grothendieck connection}
\label{subsec:Grothendieck}
As it was shown \cite{GK,B,CF3,BCM,CMW4}, one can define a flat connection $D$ on $\Hat{\Sym}(T^*M)$ with the property that $D\sigma = 0$ if and only if $\sigma = \T\varphi^*f$ for some $f\in C^\infty(M)$. Namely, $D = \dd_x + L_R$ where $R \in \Gamma(T^*M \otimes TM \otimes \Hat{\Sym}(T^*M))=\Omega^1(M,\Der(\Hat{\Sym}(T^*M)))$ is a 1-form with values in derivations of $\Hat{\Sym}(T^*M)$, which we identify with $\Gamma(TM \otimes \Hat{\Sym}(T^*M))$. We have denoted by $\dd_x$ the de Rham differential on $M$ and by $L$ the Lie derivative. 
In coordinates we have 
\begin{equation}
    R(\sigma)_\ell=-\frac{\de\sigma}{\de y^j}\left(\left(\frac{\de\varphi}{\de y}\right)^{-1}\right)_k^j\frac{\de\varphi^k}{\de x^\ell}.
\end{equation}
Define $R(x,y):=R_\ell(x,y)\dd x^\ell$, $R_\ell(x,y):=R_\ell^j(x,y)\frac{\de}{\de y^j}$, $R^j(x,y):=R^j_\ell(x,y)\dd x^\ell$, and 
\begin{equation}
    R_\ell^j=-\left(\left(\frac{\de\varphi}{\de y}\right)^{-1}\right)_k^j\frac{\de\varphi^k}{\de x^\ell}=-\delta_\ell^j+O(y).
\end{equation}


For $\sigma \in \Gamma(\Hat{\Sym}(T^*M))$, $L_R\sigma$ is given by the Taylor expansion (in the $y$ coordinates) of $$-\dd_y\sigma \circ (\dd_y\varphi)^{-1} \circ \dd_x\varphi \colon \Gamma(TM) \longrightarrow \Gamma(\Hat{\Sym}(T^*M)),$$
where we denote by $\dd_y$ the de Rham differential on the fiber.
This shows that $R$ does not depend on the choice of coordinates. One can generalize this also for any fixed vector $\xi = \xi^{i}(x) \frac{\partial}{\partial x^i}\in T_xM$ by 
\begin{equation} 
\label{eq:Aconnection_G}
D^{\xi} = \xi + \Hat{\xi}=\iota_{\xi}D,
\end{equation}
where 
\begin{equation} 
\label{eq:formal_vector}
\Hat{\xi}(x,y) = \iota_{\xi}R(x,y) = \xi^i(x)R_\ell^j(x,y)\frac{\partial}{\partial y^j}. 
\end{equation}
Here $\xi(x)$ would replace the 1-form part $\dd x^{i}$.
The connection $D$ is called the \emph{\textbf{Grothendieck connection}}. Note that its flatness is equivalent to the Maurer--Cartan equation
\begin{equation}
\dd_xR + \frac{1}{2}[R,R] = 0.
\end{equation}
Moreover, using the Poincar\'e lemma on $T_xM$ it can be shown that its cohomology is concentrated in degree 0 and is given by 
\begin{equation}
 H^0_{D}(\Gamma(\Hat{\Sym}(T^*M)))=\mathsf{T}\varphi^*C^\infty(M)\cong C^{\infty}(M).  
\end{equation}

\subsection{Lifting formal exponential maps to cotangent bundles}
\label{subsec:lifting_exp_map_cotangent_bundles}
We want to consider the case were our manifold is given by a cotangent bundle. 
\begin{prop}
\label{prop:linear_R}
If the base manifold is given by a cotangent bundle $T^*M$, the vector field $\bar R$, defined by the lift of the formal exponential map, is linear in the fiber coordinate of $T_{(q,p)}T^*M$ for any $(q,p)\in T^*M$.
\end{prop}
\begin{proof}
Let $M$ be a smooth manifold and consider a formal exponential map 
$\varphi\colon TM\longrightarrow M$. 
Moreover, let $\bar\varphi\colon TT^*M\longrightarrow T^*M$ be the lift of the formal exponential map to the cotangent bundle of $M$. Explicitly, for $(q,p)\in T^*M$, we have
$$\bar\varphi_{(q,p)}\colon T_{(q,p)}T^*M\cong T_qT_q^*M\oplus T_pT_q^*M\cong T_qM\oplus T_q^*M\longrightarrow T^*M.$$
Let $(\bar q,\bar p)\in T_{(q,p)}T^*M$, and hence $\bar q\in T_qM$ and $\bar p\in T_q^*M$. 
Note that $\varphi_q\colon T_qM\longrightarrow M$, and thus 
$$\left(\dd_{\bar q}(\varphi_{q})\right)^{*,-1}\colon T_q^*M\longrightarrow T^*_{\varphi_q(\bar q)}M,$$
since $\dd_{\bar q}(\varphi_q)\colon T_{\bar q}T_qM\cong T_qM\longrightarrow T_{\varphi_q(\bar q)}M$. Then we can write the lift of the exponential map as 
$$\bar\varphi_{(q,p)}(\bar q,\bar p)=\left(\varphi_q(\bar q),\left(\dd_{\bar q}(\varphi_{q})\right)^{*,-1}\bar p\right)\in T^*M.$$
For $x=(q,p)\in T^*M$ and $y=(\bar q,\bar p)\in T_xT^*M=T_{(q,p)}T^*M$, we want to compute 
$$\left(\dd_y(\bar\varphi_x)\right)^{-1},\qquad\dd_x(\bar\varphi_x).$$
We write $\bar\varphi^{\bar q}:=\varphi_{q}(\bar q)$ and $\bar\varphi^{\bar p}:=\left(\dd_{\bar q}(\varphi_{q})\right)^{*,-1}\bar p$. Hence we get
\begin{equation}
\dd_y\bar\varphi_x=\begin{pmatrix}\frac{\partial \bar\varphi^{\bar q}}{\partial \bar q}&\frac{\partial \bar \varphi^{\bar q}}{\partial \bar p}\\ \frac{\partial \bar \varphi^{\bar p}}{\partial \bar q}& \frac{\partial \bar\varphi^{\bar p}}{\partial \bar p}\end{pmatrix}=\begin{pmatrix}
\frac{\de \bar\varphi^{\bar q}}{\de \bar q}& \boldsymbol{0}\\ \frac{\de\bar\varphi^{\bar p}}{\de\bar q}& \frac{\de\bar\varphi^{\bar p}}{\de\bar p}
\end{pmatrix},\qquad
\dd_x\bar\varphi_x=\begin{pmatrix}\frac{\partial \bar\varphi^{\bar q}}{\partial q}& \frac{\partial \bar\varphi^{\bar p}}{\partial p}\\ \frac{\partial \bar\varphi^{\bar p}}{\partial q}&\frac{\partial \bar\varphi^{\bar p}}{\partial p}\end{pmatrix}=\begin{pmatrix}\frac{\partial \bar\varphi^{\bar q}}{\partial q}&\boldsymbol{0}\\ \frac{\partial\bar\varphi^{\bar p}}{\partial q}&\boldsymbol{0}\end{pmatrix}.
\end{equation}
Moreover, we have 
\begin{equation}
    \left(\dd_y\bar\varphi_x\right)^{-1}=\begin{pmatrix}\left(\frac{\de\bar\varphi^{\bar q}}{\de\bar q}\right)^{-1}& \boldsymbol{0}\\ -\left(\frac{\de\bar\varphi^{\bar p}}{\de\bar p}\right)^{-1}\circ \left(\frac{\de\bar\varphi^{\bar p}}{\de\bar q}\right)\circ \left(\frac{\de\bar\varphi^{\bar q}}{\de\bar q}\right)^{-1}&\left(\frac{\de\bar\varphi^{\bar p}}{\de\bar p}\right)^{-1}\end{pmatrix}.
\end{equation}
Thus, we get
\begin{equation}
-\left(\dd_y\bar\varphi_x\right)^{-1}\circ \dd_x\bar\varphi_x=\begin{pmatrix}-\left(\frac{\de\bar\varphi^{\bar q}}{\de\bar q}\right)^{-1}\circ \left(\frac{\de\bar\varphi^{\bar q}}{\de q}\right)&\boldsymbol{0}\\ \left(\frac{\de\bar\varphi^{\bar p}}{\de\bar p}\right)^{-1}\circ \left(\frac{\de \bar\varphi^{\bar p}}{\de \bar q}\right)\circ \left(\frac{\de \bar\varphi^{\bar q}}{\de\bar q}\right)^{-1}\circ \left(\frac{\de \bar\varphi^{\bar q}}{\de q}\right)-\left(\frac{\de\bar\varphi^{\bar p}}{\de\bar p}\right)^{-1}\circ \left(\frac{\de\bar\varphi^{\bar p}}{\de q}\right)&\boldsymbol{0}\end{pmatrix}.
\end{equation}
If we consider the lift $\bar R=-\dd_y\circ \left(\dd_y\bar\varphi_x\right)^{-1}\circ\dd_x\bar\varphi_x$, we get that $\bar R$ is linear in $\bar p$ as claimed. 
\end{proof}
\begin{rem}
Proposition \ref{prop:linear_R} will simplify the graphs in the Feynman graph expansion of the formal global Poisson Sigma Model for cotangent targets as we will see later on.
\end{rem}

\subsection{The deformed Grothendieck connection}
Let $M\subset \R^d$ be an open subset and consider a Poisson structure $\pi$ on $\R^d$. We will denote its associated Weyl bundle\footnote{Typically, one only speaks of a Weyl bundle if the underlying manifold $M$ is symplectic. However, the construction is general for any manifold $M$.} by $\calW:=\Hat{\Sym}(T^*M)[\![\hbar]\!]$ similarly as in the symplectic case.
Using Kontsevich's formality map, one can construct a global connection $\calD$ on $\Gamma(\calW)$ as follows: For some vector field $\xi$, define a differential operator
\begin{equation}
A(\xi,\pi):=\sum_{j=1}^\infty\frac{\hbar^j}{j!}\calU_{j+1}(\xi,\pi,\ldots,\pi)\in \calD_{poly}^0(M),
\end{equation}
using the formality map $\calU$. Define the quantized version $\calD$ of $D$ by replacing $\Hat{\xi}$ by $A(\Hat{\xi},\mathsf{T}\varphi_x^*\pi)$ in \eqref{eq:Aconnection_G}, where we consider a fixed vector $\xi\in T_xM$. Hence we have
\begin{equation}
\calD^\xi=\xi+A(\Hat{\xi},\mathsf{T}\varphi_x^*\pi).
\end{equation}
This connection can be extended to a well-defined global connection $\calD$ on $\calW$. It is in fact given as a deformation of $D$, i.e. $\calD=D+O(\hbar)$.
Moreover, $\calD$ is not flat but one can check that it is an inner derivation as in \eqref{eq:curv_fedosov}. Let $\star$ denote Kontsevich's star product. For any section $\sigma\in\Gamma(\calW)$ we have
\begin{equation}
\calD^2\sigma=[F,\sigma]_\star:=F\star\sigma-\sigma\star F,
\end{equation}
where $F\in \Omega^2(M,\calW)$ denotes the Weyl curvature tensor of $\calD$, which can be also expressed by Kontsevich's $L_\infty$-morphism. For two vector fields $\xi,\zeta$, define a function 
\begin{equation}
\label{eq:curvature_G}
F_0(\xi,\zeta,\pi):=\sum_{j=1}^\infty\frac{\hbar^j}{j!}\calU_{j+2}(\xi,\zeta,\pi,\ldots,\pi)\in \calD_{poly}^0(M)\cong C^\infty(M),
\end{equation}
in terms of the $L_\infty$-morphism $\calU$.
Then we can define the Weyl curvature tensor of $\calD$ to be given by 
\begin{equation}
\label{eq:Weyl_curv_tensor_G}
F(\xi,\zeta):=F_0(\xi,\zeta,\mathsf{T}\varphi_x^*\pi). 
\end{equation}
Moreover, one can check that the Bianchi identity $\calD F=0$ holds and that for any $\gamma\in\Omega^1(M,\calW)$ the map
\begin{equation}
\bar\calD:= \calD+[\gamma,\enspace]_\star
\end{equation}
is a derivation, i.e. $\bar\calD(\sigma\star\tau)=\bar\calD(\sigma)\star\tau+\sigma\star\bar\calD(\tau)$ for all $\sigma,\tau\in\Gamma(\calW)$. 
Computing $\bar\calD^2$ directly, one can see that the Weyl curvature tensor $\bar F$ of $\bar \calD$ is given by
\begin{equation}
\label{eq:general_fedosov}
\bar F=F+\calD\gamma+\gamma\star\gamma.
\end{equation}

\begin{prop}
\label{prop:general_fedosov_2}
There exists a $\gamma\in\Omega^1(M,\calW)$ such that $\bar F=0$. More generally, for any $\omega_\hbar=\omega_0+\hbar\omega_1+\hbar^2\omega_2+\dotsm\in \Omega^2(M,\calW)$ with $\calD\omega_\hbar=0$ and $[\omega_\hbar,\enspace]_\star=0$, there exists a $\gamma\in\Omega^1(M,\calW)$ such that 
\begin{equation}
\label{eq:general_fedosov2}
\bar F=F+\calD\gamma+\gamma\star\gamma=\omega_\hbar.
\end{equation}
\end{prop}
It is clear that \eqref{eq:general_fedosov} is the special case of \eqref{eq:general_fedosov2} for $\omega_\hbar=0$.
Proposition \ref{prop:general_fedosov_2} can be shown by using techniques of homological perturbation theory. Note that the Bianchi identity for $\bar F$ implies that $\bar\calD\omega_\hbar=\calD\omega_\hbar=0$ if $\omega_\hbar$ is a central element of the Weyl algebra endowed with Kontsevich's star product.
Equation \eqref{eq:general_fedosov2} can be seen as a more general version of \eqref{eq:fedosov} for Poisson manifolds, where $\calD$ takes the place of the symplectic connection $\nabla$ and $\bar\calD$ the one of $\bar\nabla$. 
We will say that a connection is \emph{\textbf{compatible}} if its extension to differential forms $\Omega^\bullet(M,\calW)$ is a derivation of degree $+1$ with respect to the star product on the Weyl algebra.
A compatible connection on $\Gamma(\calW)$ is called a \emph{\textbf{Fedosov connection}} if it is an inner derivation with respect to its Weyl curvature tensor and it satisfies the Bianchi identity. By the constructions above, the deformed Grothendieck connection $\calD$ is a Fedosov connection as well as any symplectic connection $\nabla$ on the tangent bundle of a symplectic manifold as in Fedosov's construction. Note that, as we have seen, if $\calD$ is a Fedosov connection, then $\bar\calD=\calD+[\gamma,\enspace]_\star$ is also a Fedosov connection.

\subsection{Grothendieck connection on symplectic manifolds}
\label{subsec:Grothendieck_conn_symplectic_manifolds}
Let $(M,\omega)$ be a symplectic manifold which can be considered as a special case of a Poisson manifold with Poisson structure $\pi$ coming from the symplectic form. By Darboux's theorem, we consider a constant symplectic form $\omega^\varphi:=\varphi^*_x\omega$ lifted to the formal construction for any $x\in M$. Note that in this case $R$ is a 1-form on $M$ with values in formal Hamiltonian vector fields\footnote{Recall that for a function $f$ one can construct a unique vector field $X_f$, such that $\iota_{X_f}\omega=-\dd f$; the vector field $X_f$ is called the Hamiltonian vector field of $f$ and $f$ is called the Hamiltonian function of $X_f$.} for the corresponding Hamiltonian functions $h_x$ such that $h_x|_{y=0}=0$. 
For any $x\in M$, $h_x$ is a 1-form with values in $\Hat{\Sym}(T^*M)$ and for any section $\sigma\in\Gamma(\calW)$ we get $\calD\sigma_x=\dd_x\sigma_x+\frac{1}{2\hbar}[h_x,\sigma_x]_\star$. For the Weyl curvature tensor we get $F_x(\xi,\zeta)=\frac{1}{4\hbar^2}([\langle h_x,\xi\rangle,\langle h_x,\zeta\rangle]_\star-2\hbar\{\langle h_x,\zeta\rangle,\langle h_x,\zeta\rangle\})$. Consider a symplectic connection $\nabla$ on $TM$, which induces a connection $\calD$ on $\Gamma(\calW)$, which acts as a derivation on the Weyl algebra. Its curvature is then given by $\calD^2=\frac{1}{2\hbar}[F,\enspace]$ with $F\in\Omega^2(M,\calW)$ given by $F=-\frac{1}{2}\omega(\nabla^2y,y)$ is the quadratic form on $TM$ associated to the curvature $\nabla^2$ of $\nabla$. Fedosov showed that for a closed 2-form $\omega_\hbar=-\frac{1}{2\hbar}\omega+\omega_0+\hbar\omega_1+\hbar^2\omega_2+\dotsm\in \Omega^2(M,\R)[\![ \hbar ]\!]$ the equation 
\begin{equation}
\nabla^2+\nabla\gamma_\hbar+\gamma_\hbar\star\gamma_\hbar=\omega_\hbar
\end{equation}
has a solution $\gamma_\hbar=-\frac{1}{2\hbar}\omega_{ij}y^{i}\dd x^j+\gamma_0+\hbar\gamma_1+\hbar^2\gamma_2+\dotsm\in \Omega^1(M,\calW)$ such that $\gamma_\hbar|_{y=0}=0$. Moreover, we consider the formal exponential map coming from the symplectic connection $\nabla$
\begin{equation}
\varphi_x^{i}(y)=x^i+y^{i}+\frac{1}{2}\sum_{k,\ell}\Gamma_{k\ell}^{i}y^ky^\ell+\dotsm
\end{equation}
Then the connection $\nabla+[\gamma_\hbar,\enspace]_\star$ is given by $\bar\calD=\calD+[\gamma,\enspace]_\star$ with $\gamma_\hbar=-\frac{1}{2\hbar}h_x+\gamma$, where $\gamma$ is a solution of \eqref{eq:general_fedosov2} with $\omega_\hbar=\omega_0+\hbar\omega_1+\hbar^2\omega_2+\dotsm$.
The star product constructed in this way, using a closed two form $\omega_\hbar\in \Omega^2(M,\R)[\![ \hbar ]\!]$, is equivalent to the one constructed by Fedosov associated to the class $-\frac{1}{2\hbar}\omega+\omega_\hbar$. Note that the deformations of the symplectic form are in one-to-one correspondence with their characteristic classes, which are formal power series $\omega_\hbar=\omega_0+\hbar\omega_1+\hbar^2\omega_2+\dotsm$, with $\omega_i\in H^2(M,\R)$ such that $-\omega_0$ is the class of the symplectic form $\omega$. For more details on these constructions see \cite{CattaneoFelderTomassini2002}.

\subsection{Globalization of Kontsevich's star product}
\label{subsec:globalization}
Consider again a Poisson manifold $(M,\pi)$. As already mentioned, the algebra of smooth functions on $M$ is isomorphic to the subalgebra of $D$-closed sections of $\Hat{\Sym}(T^*M)$. Denote by $H^0_{\bar\calD}(\Gamma(\calW))$ the subalgebra of $\Gamma(\calW)$ consisting of $\bar\calD$-closed sections of $\calW$. Since $D$ and $\bar\calD$ are flat connections we have natural cochain complexes $(\Gamma(\Hat{\Sym}(T^*M)),D)$ and $(\Gamma(\calW),\bar\calD)$.
\begin{prop}
The subalgebra $H^0_{\bar\calD}(\Gamma(\calW))$ provides a deformation quantization of $(M,\pi)$.
\end{prop}
More precisely, we can construct a \emph{\textbf{cochain map}} 
\begin{equation}
\label{eq:cochain_map}
\rho\colon (\Gamma(\Hat{\Sym}(T^*M),D)\longrightarrow (\Gamma(\calW),\bar\calD),
\end{equation}
which implies a \emph{\textbf{quantization map}}
\begin{equation}
    \rho\colon C^\infty(M)\cong H^0_D(\Gamma(\Hat{\Sym}(T^*M)))\longrightarrow H^0_{\bar\calD}(\Gamma(\calW)).
\end{equation}
This map induces an isomorphism $C^\infty(M)[\![ \hbar ]\!]\xrightarrow{\sim}H^0_{\bar\calD}(\Gamma(\calW))$, since there are no cohomological obstructions. Note that this is the analogue of the symbol map as in Fedosov's quantization. Moreover, there is a unique $\rho$ for each $\bar\calD$ such that $\rho|_{y=0}=\id$. Using this map, one can define a global version of Kontsevich's star product, defined on the whole Poisson manifold $M$ by
\begin{equation}
f \star_M g:=[\rho^{-1}(\rho(\mathsf{T}\varphi^* f)\star\rho(\mathsf{T}\varphi^*g))]\Big|_{y=0}.
\end{equation}
Indeed, the map $\rho$ sends $D$-flat sections to $\bar\calD$-flat sections since $\rho$ is a cochain map, i.e. we have $\rho\circ D=\bar\calD\circ\rho$, and by compatibility with the star product, one can obtain that $J:=\rho(\mathsf{T}\varphi^* f)\star \rho(\mathsf{T}\varphi^*g)$ is again $\bar\calD$-closed because $\mathsf{T}\varphi^*f$ is $D$-closed for all $f\in C^\infty(M)[\![ \hbar ]\!]$. But since $J$ is $\bar\calD$-closed, we know that it has to lie in the image of $\rho$. Hence there exists some $j\in \Gamma(\Hat{\Sym}(T^*M))$ such that $\rho(j)=J$. This implies that $j$ is $D$-closed and thus of the form $j=\mathsf{T}\varphi^*\tilde j$ for some $\tilde j\in C^\infty(M)[\![ \hbar ]\!]$. Setting the formal variables $y=0$ one finds a global construction for the star product.

This approach generalizes Fedosov's construction for the Moyal product, to the globalization of Kontsevich's star product. It can be translated into field theoretic concepts using the Grothendieck connection together with the Poisson Sigma Model as we will also briefly recall in Section \ref{subsec:formal_global_action}.

\section{The Poisson Sigma Model and its globalization}

\subsection{The classical model}
\begin{figure}[!ht]
\centering
\tikzset{
particle/.style={thick,draw=black},
particle2/.style={thick,draw=black, postaction={decorate},
    decoration={markings,mark=at position .9 with {\arrow[black]{triangle 45}}}},
gluon/.style={decorate, draw=black,
    decoration={coil,aspect=0}}
 }
\begin{tikzpicture}[x=0.05\textwidth, y=0.05\textwidth]
\node[](1) at (0,0){};
\node[](2) at (5,0){};
\node[](3) at (4,-2){};
\node[](4) at (4.4,-2.3){$1$};
\draw[fill=black] (3) circle (.07cm);
\node[](5) at (1,-2){};
\node[](6) at (0.6,-2.3){$0$};
\draw[fill=black] (5) circle (.07cm);
\node[](7) at (2.5,2.5){};
\node[](8) at (2.5,2.9){$\infty$};
\draw[fill=black] (7) circle (.07cm);
\node[](9) at (2.5,0){$\mathbb{D}$};
\node[](2) at (5,0){};
\semiloop[particle]{1}{2}{0};
\semiloop[particle]{2}{1}{180};
\end{tikzpicture}
\caption{Cyclically ordered points on $S^1=\partial \mathbb{D}$}
\label{fig:cyc}
\end{figure}
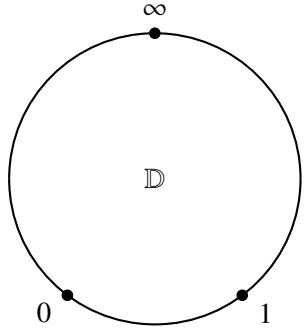

The data for the \emph{\textbf{Poisson Sigma Model}} consists of a Poisson manifold $(M,\pi)$, a compact, connected $2$-manifold $\Sigma$ (possibly with boundary), a map $X\colon \Sigma\longrightarrow M$, a 1-form $\eta\in \Gamma(\Sigma,T^*\Sigma\otimes X^*T^*M)=\Omega^1(\Sigma,X^*T^*M)$, and an action functional 
\begin{equation}
\label{eq:PSM_classical}
S_\Sigma(X,\eta)=\int_\Sigma\left(\langle \eta,\dd X\rangle+\frac{1}{2}\langle \pi(X),\eta\land\eta\rangle\right).
\end{equation}
We consider the space of fields as vector bundle maps $F_\Sigma=\Map_{\textnormal{VecBun}}(T\Sigma,T^*M)$, i.e. we have the following diagram
\[
\begin{tikzcd}
T\Sigma\arrow[r,""]\arrow[dr,swap,"\pi_\Sigma"]\arrow[bend left]{rr}{(X,\eta)}&X^*T^*M \arrow[r,""] \arrow[d, swap,"\pi_1"]
& T^*M \arrow[d, "\pi_2"] \\
T^*\Sigma\otimes X^*T^*M\arrow[bend right,swap]{r}{\pi_\otimes}&\arrow[l,"\eta"]\Sigma \arrow[r, "X"]
& M
\end{tikzcd}
\]
Consider now the case where $\de\Sigma\not=\varnothing$ and let $\iota_{\de\Sigma}\colon \de\Sigma\hookrightarrow \Sigma$ denote the inclusion of the boundary. Then we set the boundary conditions such that $\iota^*_{\de\Sigma}\eta=0$. This is convenient to choose, since the Euler--Lagrange equations are given by
\begin{equation}
\dd X^{i}+\pi^{ij}(X)\eta_j=0,\qquad \dd\eta_i+\frac{1}{2}\de_i\pi^{jk}(X)\eta_j\land\eta_k=0.
\end{equation}
Hence, it is easy to consider the solution where $X=const.$ and $\eta=0$.
In \cite{CF1} it was shown that this model is directly connected to Kontsevich's star product as formulating it by a quantum field theory where the space-time manifold $\Sigma$ is modelled by the disk $\mathbb{D}=\{x\in \R^2\mid \|x\|\leq1\}$. If we choose three points $0,1,\infty$ on the boundary $\de \mathbb{D}$ counterclockwise (i.e. if we move from $0$ counterclockwise on the boundary, we will first meet $1$ and then $\infty$, see Figure \ref{fig:cyc}), Kontsevich's star product is given by the semiclassical expansion of the \emph{\textbf{path integral}} modelled by the Poisson Sigma Model as
\begin{equation}
\label{eq:path_int_kontsevich}
f\star g(x)=\int_{X(\infty)=x}f(X(1))g(X(0))\exp\left(\frac{\I}{\hbar}S_{\mathbb{D}}(X,\eta)\right).
\end{equation}

\subsection{BV formulation}
The \emph{\textbf{Batalin--Vilkovisky (BV) formalism}} \cite{BV1,BV2,BV3} is a way of dealing with gauge theories\footnote{In fact, the Poisson Sigma Model is a nontrivial example of a gauge theory where the BV formalism is actually needed for quantization, showing the importance of the formalism.}, i.e. of theories where the action is invariant under certain symmetries.
There we usually associate to a space-time manifold $\Sigma$ a BV space of fields $\calF_\Sigma$ (in general, if one starts with the BRST formalism, we get $\calF_{\textnormal{BV}}=T^*[-1]\calF_{\textnormal{BRST}}$), which is a $\mathbb{Z}$-graded \emph{\textbf{supermanifold}}, endowed with a $(-1)$-shifted symplectic structure $\omega_\Sigma$ and an action functional $S_\Sigma\in \calO(\calF_\Sigma)$ of degree 0 such that $\{S_\Sigma,S_\Sigma\}=0$ (\emph{\textbf{Classical Master Equation}}), where $\{\enspace,\enspace\}$ denotes the \emph{\textbf{BV bracket}} coming from the odd symplectic form $\omega_\Sigma$. Here we denote by $\calO(X)$ functions on a space $X$.
We would like our theory to be local, i.e. we require the action to be given as an integral over some Lagrangian density $\mathscr{L}$ depending on fields and higher derivatives
\begin{equation}
S(\phi)=\int_\Sigma \mathscr{L}(\phi,\de\phi,\ldots), \quad \phi\in\calF_\Sigma.
\end{equation}
Moreover, we consider the \emph{\textbf{BV Laplacian}} $\Delta$, acting on functions on $\calF_\Sigma$. We will denote by $\calO_{\textnormal{loc}}(\calF_\Sigma)$ the space of such local functions on $\calF_\Sigma$. One can check that $(\calO_{\textnormal{loc}}(\calF_\Sigma),\Delta)$ is a \emph{\textbf{BV algebra}} (see Appendix \ref{app_sec:BV_algebras})\footnote{One should be aware that this construction is only formal, since the discussion in Appendix \ref{app_sec:BV_algebras} only applies to finite-dimensional manifolds.}. Moreover, we can define a cohomological vector field (similarly as in the linear case, which would be the usual BRST charge) as the degree $+1$ Hamiltonian vector field $Q_\Sigma$ of $S_\Sigma$, i.e. $\iota_{Q_\Sigma}\omega_\Sigma=-\dd_{\calF_\Sigma}S_\Sigma$. Then we have $[Q_\Sigma,Q_\Sigma]=0$ and $Q_\Sigma=\{S_\Sigma,\enspace\}$. Here $\dd_{\calF_\Sigma}$ denotes the de Rham differential on the BV space of fields $\calF_\Sigma$.

\subsubsection{BV formulation for the Poisson Sigma Model}
Let everything be as in the setting of the Poisson Sigma Model. The BV space of fields is given by $\calF_\Sigma=\Map_{\text{SupMnf}}(T[1]\Sigma,T^*[1]M)$ which are maps between supermanifolds, where for the superfields $(\mathsf{X},\boldsymbol{\eta})\in\calF_\Sigma$ we have the \emph{\textbf{BV action functional}}
\begin{equation}
S_{\Sigma}(\mathsf{X},\boldsymbol{\eta})=\int_{T[1]\Sigma}\left(\langle\boldsymbol{\eta},\boldsymbol{D}\mathsf{X}\rangle+\frac{1}{2}\langle\pi(\mathsf{X}),\boldsymbol{\eta}\land\boldsymbol{\eta}\rangle\right),
\end{equation}
where $\boldsymbol{D}=\theta^\mu\frac{\de}{\de x^\mu}$ is the superdifferential for even coordinates $(x^\mu)$ and odd coordinates $(\theta^\mu)$ and $\langle\enspace,\enspace\rangle$ denotes the pairing of tangent and cotangent space of $M$. One can write out the components of the superfields in terms of fields, antifields and ghosts as follows
\begin{align}
\mathsf{X}^{i}&=X^{i}+\eta^{+,i}_\mu\theta^{\mu}+\frac{1}{2}\beta^{+,i}_{\mu\nu}\theta^{\mu}\theta^{\nu},\\
\boldsymbol{\eta}_i&=\beta_{i}+\eta_{i,\mu}\theta^{\mu}+\frac{1}{2}X^+_{i,\mu\nu}\theta^{\mu}\theta^{\nu},
\end{align}
where $\beta$ denotes the ghost field. For a field $\phi$ we denote by $\phi^+$ its antifield. Note that we have the relation $\gh(\phi)+\gh(\phi^+)=-1$ and $\deg(\phi)+\deg(\phi^+)=2$, where ``$\gh$'' denotes the \emph{\textbf{ghost number}} which corresponds to the $\Z$-grading on $\calF_\Sigma$, and ``$\deg$'' denotes the form degree. Thus we get 
\begin{align*}
    \deg(X)&=0,\quad \deg(X^+)=2,\quad \gh(X)=0,\quad \gh(X^+)=-1\\
    \deg(\eta)&=1,\quad \deg(\eta^+)=1,\quad \gh(\eta)=0,\quad \gh(\eta^+)=-1\\
    \deg(\beta)&=0,\quad \deg(\beta^+)=2,\quad \gh(\beta)=1,\quad \gh(\beta^+)=-2
\end{align*}
In local coordinates we have
\begin{equation}
S_{\Sigma}(\mathsf{X},\boldsymbol{\eta})=\int_\Sigma\left(\boldsymbol{\eta}_i\land\dd\mathsf{X}^{i}+\frac{1}{2}\pi^{ij}(\mathsf{X})\boldsymbol{\eta}_i\land\boldsymbol{\eta}_j\right),
\end{equation}
where now $\dd$ denotes the de Rham differential on $\Sigma$.
Note that the BV action has the same form as the classical action \eqref{eq:PSM_classical} and thus it produces the same Euler--Lagrange equations, where the classical fields are replaced by the superfields and the de Rham differential $\dd$ on $\Sigma$ is replaced by the superdifferential $\boldsymbol{D}$.

\subsubsection{Equivariant BV formulation}
Consider a Lie algebra $\mathfrak{g}$ acting on $\Sigma$ via a vector field $v_X$ for some $X\in \mathfrak{g}$. Note that the cohomological vector field is given by
\begin{equation}
    Q_\Sigma = \dd_{\calF_\Sigma}+\Hat{\Theta}_\pi=\{S_\Sigma,\enspace\},
\end{equation}
where $\dd_{\calF_\Sigma}$ and $\Hat{\Theta}_\pi$ are the Hamiltonian vector fields for the Hamiltonians 
\begin{equation}
    S_0=\int_\Sigma\langle \boldsymbol{\eta},\dd\mathsf{X}\rangle\quad\text{and}\quad S_\pi=\frac{1}{2}\int_\Sigma\langle\pi(\mathsf{X}),\boldsymbol{\eta}\land \boldsymbol{\eta}\rangle
\end{equation}
respectively. Then one can check that the Classical Master Equation $Q_\Sigma(S_\Sigma)=\{S_\Sigma,S_\Sigma\}=0$ holds. Consider some variable $u$ of cohomological degree $2$ and define a $\mathfrak{g}$-DG algebra $\calO(\calF_\Sigma)[u]:=\calO(\calF_\Sigma)\otimes \Sym(\mathfrak{g}^*)$. We can define the equivariant extension of the BV action in the Cartan model as
\begin{equation}
    \label{eq:equivariant_BV_action}
    S_\Sigma^{c}=S_\Sigma+uS_{\iota_{v_X}},
\end{equation}
for $X\in\mathfrak{g}$. Choosing a basis $(e_j)$ of $\mathfrak{g}$, we get 
\begin{equation}
    S_\Sigma^{c}=S_\Sigma+u^jS_{\Hat{\iota}_{v_j}}, 
\end{equation}
where $S_{\Hat{\iota}_{v_j}}$ is the Hamiltonian of $\Hat{\iota}_{v_j}$ which is the vector field on $\calF_\Sigma$ obtained from the vector field $\iota_{v_j}$, such that 
\begin{equation}
    Q_\Sigma^c=\{S_\Sigma^c,\enspace\}=\dd_{\calF_\Sigma}+\Hat{\Theta}_\pi-u^j\Hat{\iota}_{v_j},
\end{equation}
is the differential of the Cartan model of equivariant cohomology. Hence $S_\Sigma^c\in \calO_{\textnormal{loc}}(\calF_\Sigma)[u]^\mathfrak{g}$. Moreover, the Classical Master Equation extends to the \emph{\textbf{equivariant Classical Master Equation}} 
\begin{equation}
    \label{eq:equivariant_CME}
    \frac{1}{2}\{S_\Sigma^c,S^c_\Sigma\}+u^jS_{\Hat{L}_{v_j}}=0,
\end{equation}
where $S_{L_{v_j}}$ is the Hamiltonian of the vector field $\Hat{L}_{v_j}$ which is the vector field on $\calF_\Sigma$ defined by $L_{v_j}$. The \emph{\textbf{equivariant Quantum Master Equation}} is then given by 
\begin{equation}
    \label{eq:equivariant_QME}
    -u^j(S_{\Hat{L}_{v_j}}+\I\hbar\Delta S_{\Hat{\iota}_{v_j}})-\I\hbar\Delta S_\Sigma=0.
\end{equation}
For the case where $\Sigma=\mathbb{D}$ we have an $S^1$-action and hence we can consider the $S^1$-equivariant theory. For more details on the equivariant BV construction see \cite{CattZabz2019}.

\subsection{Splitting of the space of fields}
\label{subsec:splitting_of_the _space_of_fields}
We consider a \emph{\textbf{symplectic splitting}} of the space of fields into \emph{\textbf{residual fields}} (low energy fields) and \emph{\textbf{fluctuations}} (high energy fields), which, for the examples considered in this paper, exists by techniques of Hodge theory (see e.g. \cite{CMR2}). We write 
\begin{equation}
\calF_\Sigma=\calM_1\times\calM_2,
\end{equation}
where $\calM_1$ is the space of residual fields and $\calM_2$ the space of fluctuation fields.
We want to assume that $\calM_1$ is finite-dimensional, which is the case for $BF$-like theories (such as the Poisson Sigma Model). In this case it is always possible to find a split $\Delta=\Delta_1+\Delta_2$, where $\Delta_j$ is a BV Laplacian on $\calM_j$, $j=1,2$. Consider a half-density $f$ on $\calF_\Sigma$. Then for any Lagrangian submanifold $\calL\subset \calM_2$ we get
\begin{equation}
\label{eq:BV_int_1}
\Delta_1\int_\calL f=\int_\calL \Delta f.    
\end{equation}
Here $\int_\calL$ denotes the \emph{\textbf{BV pushforward}}, which is defined on half-densities by restricting the half-density to $\calL$ which makes it a density and apply the Berezinian integral. Note that the choice of $\calL$ is equivalent to gauge-fixing since, assuming the \emph{\textbf{Quantum Master Equation}} $\Delta\exp\left(\frac{\I}{\hbar}S_\Sigma\right)=0\Leftrightarrow \{S_\Sigma,S_\Sigma\}-2\I\hbar\Delta S_\Sigma=0$, we have an invariance of the BV pushforward $\int_\calL\exp\left(\frac{\I}{\hbar}S_\Sigma\right)$ under continuous deformation of $\calL$ up to $\Delta_1$-exact terms. This is due to the following theorem.
\begin{thm}[Batalin--Vilkovisky]
The following holds:
\begin{itemize}
    \item If $f=\Delta g$, then $\int_\calL f=0$,
    \item If $\Delta f=0$, then $\frac{\dd}{\dd t}\int_{\calL_t}f=0$, for a continuous family $(\calL_t)$ of Lagrangian submanifolds.
\end{itemize}
\end{thm}
If we take $f=\exp\left(\frac{\I}{\hbar}S_\Sigma\right)$, we get that the Quantum Master Equation has to hold for the second point of the theorem.
For $BF$-like theories, $\calF_\Sigma$ is given as the direct sum of two complexes $\calC\oplus \bar\calC$ endowed with the differentials $\delta$ and $\bar \delta$. We want them to be endowed with a nondegenerate pairing $\langle \enspace,\enspace\rangle$ of degree $-1$ such that the differentials are related by $\langle B,\delta A\rangle=\langle \bar \delta B,A\rangle$ for all $A\in \calC$ and $B\in \bar \calC$.
In that case $\calM_1$ is given by the cohomology $\calH\oplus \bar\calH$ and $\calM_2$ is just a complement in $\calF_\Sigma$. For the case of the Poisson Sigma Model with boundary ($\de\Sigma\not=\varnothing$) such that the boundary is given by the disjoint union of two boundary components $\de_1\Sigma$ and $\de_2\Sigma$ we have 
\begin{equation}
\label{eq:diff_forms}
\calF_\Sigma=\Omega^\bullet(\Sigma,\de_1\Sigma)\otimes T_xM\oplus \Omega^\bullet(\Sigma,\de_2\Sigma)\otimes T^*_xM[1],
\end{equation}
for a constant background field $x\colon \Sigma\longrightarrow M$, and thus 
\begin{equation}
\calM_1=H^\bullet(\Sigma,\de_1\Sigma)\otimes T_xM\oplus H^\bullet(\Sigma,\de_2\Sigma)\otimes T^*_xM[1].
\end{equation}
According to the splitting of the space of fields, we write $\mathsf{X}=\mathsf{x}+\mathscr{X}$ and $\boldsymbol{\eta}=\mathsf{e}+\mathscr{E}$, where $\mathsf{x},\mathsf{e}\in\calM_1$ and $\mathscr{X},\mathscr{E}\in\calM_2$.
\begin{rem}
Note that functions on the shifted tangent bundle $T[1]\Sigma$ are given by the algebra of differential forms $\Omega^\bullet(\Sigma)$, which indeed allows us to write the space of fields as in \eqref{eq:diff_forms}. Moreover, if we would have a manifold $\Sigma$ with boundary $\de\Sigma=\de_1\Sigma\sqcup\de_2\Sigma$ as mentioned before, we can split the space of fields as $\calF_\Sigma=\calB\times \calM_1\times \calM_2$, where $\calB$ would denote the leaf space of the symplectic foliation induced by a chosen polarization on the boundary to perform geometric quantization, where we would choose the convenient $\frac{\delta}{\delta\E}$-polarization on $\de_1\Sigma$ and the opposite $\frac{\delta}{\delta\mathbb{X}}$-polarization on $\de_2\Sigma$, where $\E$ and $\mathbb{X}$ denote the $\boldsymbol{\eta}$- and $\mathsf{X}$-boundary fields respectively (elements of the leaf space $\calB$). Moreover, one can always obtain a symplectic structure on the space of boundary fields by symplectic reduction. Hence, by techniques of geometric quantization, one would obtain a vector space for each boundary component and one can speak of ``boundary states'' as elements of these spaces. 
This construction is needed for treating the Poisson Sigma Model in the Hamiltonian approach of the \emph{\textbf{BFV formalism}} (space of boundary fields) coupled together to the BV formalism, which is called the \emph{\textbf{BV-BFV formalism}} \cite{CMR1,CMR2,CattMosh1}. We will not use the BV-BFV construction, since we will only deal with the disk $\mathbb{D}$ with one single boundary component together with the boundary condition $\iota^*_{\de \mathbb{D}}\eta=0$.
\end{rem}

\subsection{The formal global action}
\label{subsec:formal_global_action}
Let us consider for a multivector field $\xi_k\in\Gamma(\bigwedge^{k}TM)$ the local functional\footnote{Note that we are secretly using a formal exponential map $\varphi$. In particular, $\xi_k$ should be $\varphi^*\xi_k$.}
\begin{equation}
S_\xi(\mathsf{X},\boldsymbol{\eta}):=\frac{1}{k!}\int_\Sigma \xi^{i_1,\ldots,i_k}(\mathsf{X})\boldsymbol{\eta}_{i_1}\land\cdots\land\boldsymbol{\eta}_{i_k}\in \calO_{\textnormal{loc}}(\calF_\Sigma).
\end{equation}
Note that for any $k\geq0$ we have $Q_\Sigma(S_\xi)=\{S_\Sigma,S_\xi\}=0$. 
In \cite{CMW3} it was shown that the Poisson Sigma Model action can be formally globalized by adding another term to the action, which is given by
\begin{equation}
\varphi_x^*S_R(\hatX,\hateta)=\int_\Sigma R^{j}_{i}(x,\hatX)\hateta_j\land\dd x^{i},
\end{equation}
where $\hatX$ and $\hateta$ are defined by the following equations
\begin{equation}
\mathsf{X}=\varphi_x(\hatX) ,\quad \boldsymbol{\eta}=(\dd\varphi_x)^{*,-1}\hateta.
\end{equation}
Recall that $x\colon\Sigma\longrightarrow M$ denotes a constant background field. Denote by $S_0$ the free part of the action, i.e. $S_0:=\int_\Sigma \boldsymbol{\eta}_i\land\dd\mathsf{X}^{i}$. Lifting the Poisson Sigma Model action to the formal construction, we get the \emph{\textbf{formal global action}}
\begin{align}
\label{eq:formal_global_action}
\begin{split}
S^{\varphi_x}(\hatX,\hateta)&:=\varphi_x^*S_0+\mathsf{T}\varphi_x^*S_\pi+\varphi_x^*S_R\\
&=\int_\Sigma\left(\hateta_i\land\dd\Hat{\mathsf{X}}^{i}+\frac{1}{2}(\mathsf{T}\varphi_x^*\pi)^{ij}(\hatX)\hateta_i\land\hateta_j+R^{j}_{i}(x,\Hat{\mathsf{X}})\Hat{\boldsymbol{\eta}}_j\land\dd x^{i}\right).
\end{split}
\end{align}
If we denote by $\pi^\varphi:=\mathsf{T}\varphi^*\pi$, we can observe $\mathsf{T}\varphi^*S_\pi=S_{\pi^\varphi}$.
Note that the de Rham differential in $\dd \hatX^{i}$ is on $\Sigma$ and the de Rham differential in $\dd x^{i}$ is on the moduli space of constant solutions to the Euler--Lagrange equations
\begin{equation}
\calM_{cl}:=\{(X,\eta)\in \calF_\Sigma\mid X=x\colon\Sigma\longrightarrow M \textnormal{ constant map, }\eta=0\}\cong M.
\end{equation}
\begin{rem}
In general, one can consider any moduli space of solutions.
\end{rem}
One can show that \eqref{eq:formal_global_action} satisfies the \emph{\textbf{differential Classical Master Equation}}
\begin{equation}
\dd_xS^{\varphi_x}+\frac{1}{2}\{S^{\varphi_x},S^{\varphi_x}\}=0.
\end{equation}
For quantization, consider the partition function, given by \eqref{eq:formal_global_action}
\begin{equation}
Z_x=\int_\calL\exp\left(\frac{\I}{\hbar}S^{\varphi_x}\right)
\end{equation}
for some Lagrangian submanifold $\calL\subset \calM_2$. The Quantum Master equation is not satisfied in general.
It can be shown that if $\pi$ is divergence free (unimodular), the Quantum Master Equation $\Delta\exp\left(\frac{\I}{\hbar}(S_0+S_\pi)\right)=0$ holds. Another case would be if the Euler-characteristic of $\Sigma$ is zero (e.g. the torus). The choice of a unimodular Poisson structure can be seen as a renormalization procedure.
One form of renormalization is to impose that there are no tadpoles (short loops), which results in the fact that 
\begin{equation}
\Delta(\mathsf{X}(s)\boldsymbol{\eta}(s))=\sum_j(-1)^{\vert \mathsf{x}^j\vert}\mathsf{x}^j(s)\land\mathsf{e}_j(s)=:\psi(s),\quad \forall s\in\Sigma,
\end{equation}
where $\Delta$ is the BV Laplacian acting on the coefficients of the residual fields.
If we choose a volume form $\Omega$ on $M$, we can define a divergence operator $\Div_\Omega$ and thus a \emph{\textbf{renormalized BV Laplacian}} by setting (see also Appendix \ref{app_subsec:connection_to_field_theory})
\begin{equation}
\Delta S_{\xi}=\int_\Sigma \psi(\Div_\Omega\xi)^{i_1,\ldots,i_{k-1}}(\mathsf{X})\boldsymbol{\eta}_{i_1}\land\dotsm \land\boldsymbol{\eta}_{i_{k-1}}.
\end{equation}
Note that $\Delta S_\pi=0$ if $\Div_\Omega\pi=0$. 
Since $\Delta S^{\varphi_x}=0$, we get a \emph{\textbf{differential}} version of the Quantum Master Equation
\begin{equation}
\dd_x Z_x-\I\hbar\Delta Z_x=0.
\end{equation}

\begin{rem}
The formal global action has to be extended to an equivariant version such that the $S^1$-action on the disk is taken into account. 
\end{rem}

\section{Traces and algebraic index theorem}
\label{sec:traces_and_algebraic_index}
\subsection{Algebraic index theorem}
Recall that a \emph{\textbf{trace map}} on a Poisson manifold $(M,\pi)$ is a linear functional $\Tr$ on compactly supported functions $f,g\in C^\infty_c(M)$ with values in $\R(\!( \hbar )\!)$ such that 
\begin{equation}
\Tr(f\star g)=\Tr(g\star f)
\end{equation}
(hence the name ``trace"). There is a canonical trace associated to any star product coming from a symplectic manifold $(M,\omega)$ which is described within the local picture. Locally, all deformations are equivalent to the Weyl algebra and on the Weyl algebra there is a canonical trace which is constructed as an integral with respect to the Liouville measure \cite{Fedosov1994}. If we consider functions with support in neighborhoods of any point of $M$, we set the trace equal to this canonical trace restricted to these functions. 
Let $\Hat{A}(TM)$ denote the \emph{\textbf{$\Hat{A}$-genus}} of $M$, which is a characteristic class of the tangent bundle $TM$. One can express it by a de Rham representative as 
\begin{equation}
\Hat{A}(TM)=\mathrm{det}^{1/2}\left(\frac{R/2}{\sinh(R/2)}\right),
\end{equation}
where $R$ denotes the curvature of any connection on $TM$.
\begin{thm}[Nest--Tsygan \cite{NestTsygan1995}]
\label{thm:Nest-Tsygan}
Let $(M,\omega)$ be a compact symplectic manifold and let $\star$ be a star product with characteristic class $\omega_\hbar=-\omega+\hbar\omega_1+\hbar^2\omega_2+\dotsm$. Then the canonical trace associated to $\star$ obeys
\begin{equation}
\label{eq:Nest-Tsygan}
\Tr(1)=\int_M \Hat{A}(TM)\exp\left(\omega_\hbar/\hbar\right),
\end{equation}
\end{thm}
Consider again a Poisson manifold $(M,\pi)$. Let $A_\hbar:=(C^\infty(M)[\![ \hbar ]\!],\star)$ with star product coming from the Poisson structure $\pi$ (e.g. Kontsevich's star product), and denote by $CH_\bullet(A_\hbar)$ the \emph{\textbf{cyclic homology}} and by $PH_\bullet(A_\hbar)$ the \emph{\textbf{periodic cyclic homology}}. One can show that $CH_0(A_\hbar)\cong HH_0(A_\hbar)$, where $HH_\bullet(A_\hbar)$ denotes the \emph{\textbf{Hochschild homology}}. Moreover, as shown by Shoikhet and Dolgushev, the zeroth Hochschild homology is isomorphic to the zeroth \emph{\textbf{Poisson homology}}\footnote{The Poisson homology $HP_\bullet(M)$ of a Poisson manifold $(M,\pi)$ is given by the homology of the complex $(\calT_{poly}^\bullet(M),[\pi,\enspace])$, where $[\pi,\enspace]$ is the Poisson differential of degree $-1$ (note that we have taken the opposite grading).} $HP_0(M)$. If we assume that there is a volume form $\Omega$ on $M$ and that the Poisson structure $\pi$ is unimodular and $\Div_\Omega\pi=0$, we can construct a map 
\begin{align}
\label{eq:Poisson_homology_map}
\begin{split}
HP_0(M)&\longrightarrow \R,\\
f&\longmapsto \int_Mf\Omega. 
\end{split}
\end{align}
Now we can define an integration map on the zeroth periodic cyclic homology by composition 
\begin{equation}
I\colon PH_0(A_\hbar)\longrightarrow CH_0(A_\hbar)\longrightarrow HP_0(M)\xrightarrow{\int_M(-)\Omega} \R.
\end{equation}

Let $\calR$ be a DG ring with differential $\dd_\calR$. For a projective $\calR$-module $\calM$, one defines a connection to be a map
\begin{equation}
    \nabla\colon \calM\longrightarrow \calM\otimes_\calR\Omega^1_\calR,
\end{equation}
where $\Omega^1_\calR:=\calR\otimes \calR[1]$, with the usual property
\begin{equation}
    \nabla(r\otimes m)=\dd_{\calR}r\otimes m+(-1)^{\vert r\vert}r\nabla m,\quad \forall r\in\calR,\,m\in\calM.
\end{equation}
The \emph{\textbf{Atiyah class}} of a connection $\nabla$ is then defined by
\begin{equation}
    \mathrm{At}(\nabla):=[\nabla,\dd_\calR]\in\Omega^1_\calR\otimes \End_\calR(\calM).
\end{equation}
In fact, $[\mathrm{At}(\nabla)]$ measures the the obstruction to find a $\dd_\calR$-compatible connection. We define the \emph{\textbf{Chern character}} of a connection $\nabla$ by
\begin{equation}
\label{eq:Chern_character}
    \mathrm{Ch}(\nabla):=\tr\exp\left(-\frac{1}{2\pi\I}\mathrm{At}(\nabla)\right).
\end{equation}
Moreover, one can then define more generally the $\Hat{A}$-genus of a connection $\nabla$ on $\calM$ in terms of these classes by 
\begin{equation}
    \Hat{A}(\nabla)=\exp\left(-\frac{1}{2}\mathrm{Ch}(\nabla)\right)\mathrm{Td}(\nabla),
\end{equation}
where $\mathrm{Td}$ denotes the \emph{\textbf{Todd class}}, defined by 
\begin{equation}
    \mathrm{Td}(\nabla):=\frac{\mathrm{Ch}(\nabla)}{1-\exp(\mathrm{Ch}(\nabla))}.
\end{equation}

\begin{thm}[Tamarkin--Tsygan \cite{TamarkinTsygan2001}]
\label{thm:Tamarkin-Tsygan}
Let $M$ be a compact manifold with formal Poisson structure $\pi\in\hbar\Gamma(\bigwedge^2TM)[\![ \hbar ]\!]$ and $\Omega$ a volume form on $M$ with $\Div_\Omega\pi=0$ and $c\in PC_0(A_\hbar)$. Then 
\begin{equation}
\label{eq:Tamarkin-Tsygan}
I(c)=\int_M\Hat{A}_u(TM)\mathrm{Ch}(c)\exp\left(\iota_\pi/u\right)\Omega.
\end{equation}
Here 
\begin{equation}
\Hat{A}_u(TM):=\Hat{A}_0(TM)+u\Hat{A}_1(TM)+u^2\Hat{A}_2(TM)+\dotsm=\mathrm{det}^{1/2}\left(\frac{uR/2}{\sinh(uR/2)}\right),
\end{equation}
where $\Hat{A}_j(TM)\in H^{2j}(M)$ are the components of the $\Hat{A}$-genus.
\end{thm}

\subsection{A trace map for negative cyclic chains}
In \cite{CattaneoFelder2010} it was shown how one can obtain a trace map by constructing an $L_\infty$-morphism from negative cyclic chains to multivector fields with an adjunction of the formal parameter $u$ of degree $2$. Moreover, the relation to the BV formulation of the Poisson Sigma Model and how the former formula can be interpreted as an expectation value with respect to the corresponding quantum field theory was shown. However, this construction was only given for open subsets $M$ of $\R^d$. We will extend this construction to a global one using notions of formal geometry as we have seen before.

\begin{thm}[Cattaneo--Felder \cite{CattaneoFelder2010}]
Let $M$ be an open subset of $\R^d$ and consider a volume form $\Omega$ on $M$. Denote by $\delta_\Omega:=u\Div_\Omega$. Let $A=C^\infty(M)$ and let $(\calT^{-\bullet}_{poly}(M),\Div_\Omega)$ be the DG module over the DGLA $(\calT^{\bullet}_{poly}(M)[u],\delta_\Omega)$ with trivial $(\calT^{\bullet}_{poly}(M)[u],\delta_\Omega)$-action. Then there exists an $\R[u]$-linear morphism of $L_\infty$-modules over $(\calT^{\bullet}_{poly}(M)[u],\delta_\Omega)$
\begin{equation}
\calV\colon (CC_{-\bullet}^-(A),b+uB)\longrightarrow (\calT^{-\bullet}_{poly}(M)[u],u\Div_\Omega), 
\end{equation}
such that
\begin{enumerate}
    \item The zeroth Taylor component $\calV_0$ of $\calV$ vanishes on $CC^-_m(A)$, $m>0$ and for $f\in A\subset CC_0^-(A)$, $\calV_0(f)=f$.
    \item For $\xi\in \Gamma(\bigwedge^kTM)$, $\ell\geq 0$, $a=[a_0\otimes\dotsm \otimes a_m]\in CC_m^-(A)$, 
    \begin{equation}
    \calV_1(\xi v^\ell\mid a)=\begin{cases}(-1)^mu^s\xi \blacktriangleleft H(a),& \text{if $k\geq m$ and $s=k+\ell-m-1\geq 0$}\\0,&\text{otherwise}\end{cases}
    \end{equation}
    where $\blacktriangleleft\colon \calT_{poly}^k(M)\otimes \Omega^m(M,\R)\longrightarrow \calT_{poly}^{k-m}(M)$ and $H$ is the HKR map.
    \item The maps $\calV_n$ are equivariant under linear coordinate transformations and\footnote{Note that we write $\xi_1\dotsm \xi_n$ for the symmetric tensor product $\xi_1\otimes\dotsm \otimes\xi_n$ of multivector fields $\xi_1,\ldots,\xi_n\in \calT_{poly}^\bullet(M)$.} 
    \begin{equation}
    \calV_n(\xi_1\dotsm \xi_n\mid a)=\xi_1\land\calV_{n-1}(\xi_2\dotsm \xi_n\mid a)
    \end{equation}
    whenever $\xi_1=\sum (c^{i}_kx_k+d^{i})\de_i\in \calT^\bullet_{poly}(M)\subset (\calT_{poly}^\bullet(M)[u],\delta_\Omega)$ is an affine vector field and $\xi_1,\ldots,\xi_n\in (\calT_{poly}^\bullet(M)[u],\delta_\Omega)$.
\end{enumerate}
\end{thm}

The Taylor components of $\calV$ are given by maps
\begin{equation}
    \label{eq:Taylor_comp}
    \calV_n\colon (\Sym^n\calT^{\bullet+1}_{poly}(M)[u],\delta_\Omega)\otimes CC^-_{-\bullet}(A)\longrightarrow\calT^{n-1}_{poly}(M).
\end{equation}

Note that an element of degree +1 in $(\calT_{poly}^\bullet(M)[u],\delta_\Omega)$ has the form $\tilde\pi=\pi+uh$, where $\pi$ is a bivector field and $h$ a function. The Maurer--Cartan equation $\delta_\Omega\tilde\pi-\frac{1}{2}[\tilde\pi,\tilde\pi]=0$ translates to $[\pi,\pi]=0$ and
\begin{equation}
    \Div_\Omega\pi-[h,\pi]=0,
\end{equation}
and hence $\pi$ is Poisson and $h$ corresponds to the Hamiltonian function of the Hamiltonian vector field $\delta_\Omega\pi$. As we have seen, this is equivalent to the unimodularity condition.

\begin{rem}
\label{rem:PSM_morphism}
The morphism $\calV$ is in fact related to Shoikhets morphism \cite{Shoikhet2003} in the proof of Tsygan's formality theorem on chains \cite{Tsygan1999} for $M=\R^d$. It is a morphism of $L_\infty$-modules over $\calT_{poly}^{\bullet+1}(M)$ from $C_\bullet(A)$ to the DG module of differential forms $(\Omega^{-\bullet}(M,\R),\dd=0)$ and extends to \eqref{eq:cyclic_chain_formality}. The action of $\xi\in \calT_{poly}^{\bullet+1}(M)$ on $\Omega^{\bullet}(M,\R)$ is given by Lie derivative $L_\xi=\dd\circ \iota_\xi\pm \iota_\xi\circ\dd$, where the internal multiplication of vector fields is extended to multivector fields by $\iota_\xi\iota_\zeta=\iota_{\xi\land \zeta}$. This construction was globalized by Dolgushev to any manifold $M$. Moreover, recall that a volume form $\Omega\in\Omega^{d}(M,\R)$ defines an isomorphism 
\begin{align}
\begin{split}
\calT_{poly}^k(M)&\longrightarrow\Omega^{d-k}(M)\\
\xi&\longmapsto \iota_\xi\Omega, 
\end{split}
\end{align}
and thus we identify the differential $\dd$ on $\Omega^\bullet(M,\R)$ by the divergence operator $\Div_\Omega$ on $\calT^{\bullet}_{poly}(M)$. By the fact that $\calV$ is an $L_\infty$-morphism we get $\iota_{\Div \xi}\Omega=\dd \iota_\xi\Omega$.
\end{rem}
Let 
\begin{equation}
\tilde\pi_\hbar:=\hbar\pi+uh,
\end{equation}
which is a Maurer--Cartan element if $\pi+uh$ is a Maurer--Cartan element in $(\calT_{poly}^\bullet(M)[u][\![ \hbar ]\!],\delta_\Omega)$. 
If we consider the twist of $\calV$ by $\tilde\pi_\hbar$, denoted by $\calV^{\tilde\pi_\hbar}$, we can define a \emph{\textbf{trace map}} \cite{CattaneoFelder2010}
\begin{align}
\label{eq:trace_Cattaneo_Felder}
\begin{split}
\Tr\colon C^\infty_c(M)[\![h]\!]&\longrightarrow \R(\!( \hbar )\!)\\
f&\longmapsto \Tr(f)=\int_M\sum_{n=0}^\infty\frac{1}{n!}\calV_n(\tilde\pi_\hbar\dotsm\tilde\pi_\hbar\mid f)\Omega,
\end{split}
\end{align}
since $\calV^{\tilde\pi_\hbar}\colon (CC^-_{-\bullet}(A_\hbar),b+uB)\longrightarrow (\calT_{poly}^{-\bullet}(M)[u][\![ \hbar ]\!],u\Div_\Omega)$ is a chain map. We will elaborate on this fact a bit more in Section \ref{sec:main_results}. 

\begin{rem}
For a $d$-manifold $M$ denote by $V\calT_{poly}^\bullet(M):=\Omega^d(M,\bigwedge^\bullet TM)$ differential forms of degree $d$ with values in multivector fields. By the isomorphism as mentioned in Remark \ref{rem:PSM_morphism}, we can construct a natural non-degenerate pairing by
\begin{align}
\begin{split}
    \langle\enspace,\enspace\rangle\colon V\calT_{poly}^\bullet(M)\otimes \Omega_c^\bullet(M,\R)&\longrightarrow \R\\
    \xi\Omega\otimes\alpha&\longmapsto\langle\xi\Omega,\alpha\rangle:= \int_M(\iota_\xi\alpha)\Omega,
\end{split}
\end{align}
where $\xi$ is a $\bullet$-vector field and $\alpha$ is a $\bullet$-form. Here $\Omega$ denotes again a chosen volume form on $M$. We have denoted by $\Omega^\bullet_c(M,\R)$ differential forms with compact support. It is obvious that this map can be extended $u$-bilinearly. Moreover, there is an isomorphism 
\begin{align}
\begin{split}
\calT_{poly}^\bullet(M)[u]&\longrightarrow V\calT_{poly}^\bullet(M)[u]\\
\xi&\longmapsto \xi\otimes \Omega.
\end{split}
\end{align}
\end{rem}


\subsection{Construction via the Poisson Sigma Model}
\label{subsec:construction_via_PSM}
Consider now the Poisson Sigma Model on the disk . Let
\begin{equation}
Z_0:=\int_\calL \exp\left(\frac{\I}{\hbar}S_0\right),
\end{equation}
and define the \emph{\textbf{vacuum expectation value}} of an observable by the map
\begin{align}
\begin{split}
\langle\enspace\rangle_0\colon A_\hbar&\longrightarrow \R(\!( \hbar )\!)\\
f&\longmapsto \langle f\rangle_0:=\frac{1}{Z_0}\int_\calL \exp\left(\frac{\I}{\hbar}S_0\right)f.
\end{split}
\end{align}
The map $\calV_n$ can be expressed as the vacuum expectation of an observable $S_{\xi_1}\dotsm S_{\xi_j}O_{a_0,\ldots,a_m}$, where 
\begin{equation}
\label{eq:observables}
O_{a_0,\ldots,a_m}:=a_0(\mathsf{X}(t_0))\int_{t_1<t_2<\dotsm <t_m\in \de\mathbb{D}\setminus\{t_0\}}a_1(\mathsf{X}(t_1))\dotsm a_m(\mathsf{X}(t_m)).
\end{equation}
For $m$ points $t_1,\ldots,t_m\in \de\mathbb{D}$ we consider the ordering $t_0<\ldots<t_m$, which means that if we start at $t_1$ and move counterclockwise on $\de\mathbb{D}$, we wil first meet $t_2$, then $t_3$, and so on. If we embed the disk into the complex plane, i.e. we have $\mathbb{D}=\{z\in\mathbb{C}\mid \vert z\vert\leq 1\}$ and set $t_0=1$, we can express the counterclockwise condition on $\de\mathbb{D}$ by $0<\arg(t_1)<\arg(t_2)<\dotsm <\arg(t_m)<2\pi$.
The cohomology $H^\bullet(\mathbb{D})$ is $1$-dimensional and concentrated in degree zero, while the relative cohomology $H^\bullet(\mathbb{D},\de\mathbb{D})$ is $1$-dimensional and concentrated in degree two. So, for $\calM_1,\calH,\bar\calH$ as defined in Section \ref{subsec:splitting_of_the _space_of_fields}, we get $\calH=(\R^d)^*[-1]$ and $\bar\calH=\R^d$, thus $\calM_1=T^*[-1]M$. Note that functions on $\calM_1$ are then multivector fields on $M$ with reversed degree and $\Delta_1$ is given by the divergence operator $\Div_\Omega$ for the constant volume form. Note that $\Delta_1$ is an operator of degree $+1$.
For a function $f\in C^\infty(M)$ and some Lagrangian submanifold $\calL\subset \calM_2$, we have a map 
\begin{equation}
\textnormal{tr}(f):=\frac{1}{Z_0}\int_\calL \exp\left(\frac{\I}{\hbar}(S_0+S_{\tilde\pi})\right)O_f=\left\langle \exp\left(\frac{\I}{\hbar}S_{\tilde\pi}\right)O_f\right\rangle_0,
\end{equation}
given by the expectation value of the corresponding observable.
Recall that $\tilde\pi$ is a Maurer--Cartan element for a unimodular Poisson structure and that we work with the boundary condition $\iota^*_{\de\mathbb{D}}\boldsymbol{\eta}=0$. For two functions $f,g\in C^\infty(M)$, we define $O_f(\mathsf{X},\boldsymbol{\eta}):=f(\mathsf{X}(1))$ for $1\in \de\mathbb{D}$ and $O_g(\mathsf{X},\boldsymbol{\eta}):=g(\mathsf{X}(0))$ for $0\in \de\mathbb{D}$. Moreover, define\footnote{Note that this does not depend on $t$, since we fix it on the boundary.}
\begin{equation}
\textnormal{tr}_2(f,g):=\left\langle \exp\left(\frac{\I}{\hbar}S_{\tilde\pi}\right)O_{f,g}\right\rangle_0=\left\langle \exp\left(\frac{\I}{\hbar}S_{\tilde\pi}\right)f(\mathsf{X}(t))\int_{s\in\de\mathbb{D}\setminus\{t\}}g(\mathsf{X}(s))\right\rangle_0,
\end{equation}
Then we observe 
\begin{equation}
\Delta_1\textnormal{tr}_2(f,g)=\left\langle\exp\left(\frac{\I}{\hbar}S_{\tilde \pi}\right)\delta O_{f,g}\right\rangle_0,
\end{equation}
where $\delta$ was the differential on the complex $\calC$ in the definition of the space of fields in Section \ref{subsec:splitting_of_the _space_of_fields}. 
This follows from the \emph{\textbf{Ward identity}} 
\begin{equation}
    \label{eq:ward}
    \Delta_1\langle O\rangle_0=\left\langle \Delta O-\frac{\I}{\hbar}\delta O\right\rangle_0,
\end{equation}
which is true by \eqref{eq:BV_int_1}, the fact that $Z_0$ is constant on $\calM_1$ and the Leibniz rule for the BV Laplacian 
\begin{equation}
    \label{eq:BV_Laplacian_1}
    \Delta(fg)=\Delta(f)g+(-1)^{\vert f\vert}f\Delta(g)-(-1)^{\vert f\vert}\{f,g\},\quad \forall f,g\in C^\infty(M)
\end{equation}
(see also Appendix \ref{app_subsec:BV_algebras}, Equation \eqref{app:eq:BV_Laplacian_2}).
Hence by \eqref{eq:path_int_kontsevich} the two functions $f,g$ can move under the trace map from both sides\footnote{This argument follows from Stokes' theorem and the ``bubbling'' concept of the Deligne--Mumford compactification on the disk.} to each other on $\de\mathbb{D}$. Thus we get 
\begin{equation}
\label{eq:trace_condition_1}
\Delta_1\textnormal{tr}_2(f,g)=\textnormal{tr}(f\star g)-\textnormal{tr}(g\star f).
\end{equation}
Hence we get a trace on $C^\infty_c(M)$ by 
\begin{equation}
\Tr(f):=\int_M\textnormal{tr}(f)\Omega.
\end{equation}
To globalize the construction, we want to consider the formal global action $S^{\varphi}$ and additional vertices in the \emph{\textbf{Feynman graph expansion}}. In fact we will have two types of vertices in the bulk, the ones representing the formally lifted Poisson structure $\pi^\varphi_\hbar:=\mathsf{T}\varphi^*\tilde\pi_\hbar=\hbar\mathsf{T}\varphi^*\pi-v\mathsf{T}\varphi^*h$ and the ones representing the $R$ vector field coming from the definition of the Grothendick connection. We will also consider additional vertices on the boundary where we place solutions $\gamma$ of \eqref{eq:general_fedosov}. Then we can consider the vacuum expectation value 
\begin{equation}
    \left\langle \exp\left(\frac{\I}{\hbar}S^\varphi_{\pi,R}\right)O_{\rho(\mathsf{T}\varphi^*f)}\right\rangle_0,
\end{equation}
where $S^\varphi_{\pi,R}:=S_{\pi^\varphi_\hbar}+\varphi^*S_R$.
\begin{rem}
The additional vertices labeled by a solution $\gamma$ of \eqref{eq:general_fedosov} give rise to another additional term in the formal global action \cite{CMW3}. In particular, we have to consider the action
\begin{equation}
    \tilde{S}^\varphi=\varphi^*S_0+S^\varphi_{\pi,R}+\int_{\de\mathbb{D}}\hatX^*\gamma.
\end{equation}
We will call the Poisson Sigma Model with action $\tilde{S}^\varphi$ the \emph{\textbf{Fedosov-type formal global Poisson Sigma Model}} and we call $\tilde{S}^{\varphi}$ the \emph{\textbf{Fedosov-type formal global action}}.
\end{rem} 
\begin{prop}
The map
\begin{equation}
\label{eq:globalized_trace_PSM}
    \tr\colon f\longmapsto \int_M \left\langle \exp\left(\frac{\I}{\hbar}S^\varphi_{\pi,R}+\frac{\I}{\hbar}\int_{\de\mathbb{D}}\hatX^*\gamma\right)O_{\rho(\mathsf{T}\varphi^*f)}\right\rangle_0\Omega.
\end{equation}
coincides with
\begin{equation}
\label{eq:globalized_trace}
    \tr\colon f\longmapsto \sum_{n=0}^\infty\frac{1}{n!}\int_M\calV_n^{\pi^\varphi_\hbar}(R\dotsm R\mid \rho(\mathsf{T}\varphi^*f))\Omega,
\end{equation}
where we consider $\calV^{\pi^\varphi_\hbar}_n(R\dotsm R\mid \enspace)$ to be defined on the negative cyclic complex for sections of the Weyl algebra $\calW$. 
\end{prop}

This can be seen by constructing the maps $\calV_n$ in terms of graphs. We will do this in Section \ref{subsec:construction_via_graphs}.



\begin{rem}
In fact, one can construct Kontsevich's star product directly by using a path integral quantization with respect to the formal global action $S^\varphi$ as in \eqref{eq:formal_global_action}, using a similar apporach as in \cite{CF1}, with the difference that the observables on the boundary are given by $\bar\calD$-closed sections of the form $O_{\rho(\mathsf{T}\varphi^*f)}$ (see Figure \ref{fig:cyc_2}). Hence we can write it down as a path integral
\begin{equation}
    f\star_M g(x)=\rho^{-1}\left(\int_{\hatX(\infty)=x} \rho(\mathsf{T}\varphi_x^*f)(\hatX(1))\rho(\mathsf{T}\varphi_x^*g)(\hatX(0))\exp\left({\frac{\I}{\hbar}S^{\varphi_x}(\hatX,\hateta)}\right)\right)\Bigg|_{y=0}
\end{equation}

\begin{figure}[ht]
\centering
\tikzset{
particle/.style={thick,draw=black},
particle2/.style={thick,draw=black, postaction={decorate},
    decoration={markings,mark=at position .9 with {\arrow[black]{triangle 45}}}},
gluon/.style={decorate, draw=black,
    decoration={coil,aspect=0}}
 }
\begin{tikzpicture}[x=0.05\textwidth, y=0.05\textwidth]
\node[](1) at (0,0){};
\node[](2) at (5,0){};
\node[](3) at (4,-2){};
\node[](4) at (4.7,-3){$\rho(\mathsf{T}\varphi^*f)(\hatX(1))$};
\draw[fill=black] (3) circle (.07cm);
\node[](5) at (1,-2){};
\node[](6) at (0.3,-3){$\rho(\mathsf{T}\varphi^*g)(\hatX(0))$};
\draw[fill=black] (5) circle (.07cm);
\node[](7) at (2.5,2.5){};
\node[](8) at (2.5,3.2){$\rho(f\star_M g)(\hatX(\infty))$};
\draw[fill=black] (7) circle (.07cm);
\node[](9) at (2.5,0){$\mathbb{D}$};
\node[](2) at (5,0){};
\semiloop[particle]{1}{2}{0};
\semiloop[particle]{2}{1}{180};
\end{tikzpicture}
\caption{Cyclically ordered points on $S^1=\partial \mathbb{D}$}
\label{fig:cyc_2}
\end{figure}
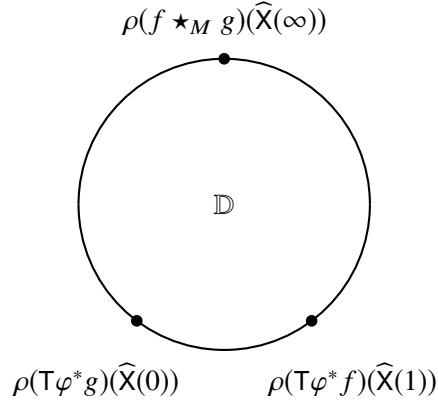

\end{rem}


\section{Feynman graphs for the globalized action}
\subsection{Construction via graphs}
\label{subsec:construction_via_graphs}
We want to describe how the Taylor components of $\calV$ are given in terms of graphs. 
In fact we have 
\begin{equation}
\label{eq:graph_repr}
\calV_n(\xi\mid a)=\sum_{\Gamma\in \calG_{\mathbf{k},m}}w_\Gamma\calV_\Gamma(\xi\mid a),
\end{equation}
where $\xi=\xi_1 \dotsm  \xi_n$, with $\xi_i\in \Gamma(\bigwedge^{k_i}TM)[u]$, $\mathbf{k}=(k_1,\ldots,k_n)$ and $a=[a_0\otimes\dotsm\otimes a_m]\in C_m(A)$. Here $w_\Gamma\in \R$ denotes the weight of a graph $\Gamma$ according to the given Feynman rules, which can be computed as integrals over configuration spaces of points on the the interior of the disk and on the boundary. We want to recall the definition of the finite set $\calG_{\mathbf{k},m}$ of oriented graphs as in \cite{CattaneoFelder2010}. 

For each graph\footnote{In addition to the rules below we will also consider it modulo graph isomorphisms which respect the partition and the orderings. The set $\calG_{\mathbf{k},m}$ is in principle given by the set of equivalence classes.} $\Gamma\in \calG_{\mathbf{k},m}$ with $n+m$ vertices ($n$ vertices in the bulk and $m$ vertices on the boundary), we assign a vertex set $V(\Gamma)=V_1(\Gamma)\sqcup V_2(\Gamma)\sqcup V_w(\Gamma)$. We will distinguish between two different types of vertices which we call the \emph{\textbf{black vertices}} $V_b(\Gamma)=V_1(\Gamma)\sqcup V_2(\Gamma)$ and the \emph{\textbf{white vertices}} $V_w(\Gamma)$. Within the black vertices we will also distinguish between vertices of \emph{\textbf{type 1}} and of \emph{\textbf{type 2}} according to the following rules.
\begin{itemize}
    \item There are $n$ vertices in $V_1(\Gamma)$. There are exactly $k_i$ edges originating at the $i$th vertex of $V_1(\Gamma)$. 
    \item There are $m$ vertices in $V_2(\Gamma)$. There are no edges originating at these vertices.
    \item There is exactly one edge pointing at each vertex in $V_w(\Gamma)$ and no edge originating from it.
    \item There are no edges starting and ending at the same vertex. 
    \item For each pair of vertices $(i,j)$ there is at most one edge from $i$ to $j$.
\end{itemize}
Each multivector field $\xi_i$ can be endowed with a power of the formal parameter $v^\ell$, which represent the residual field assigned to a black vertex.

\begin{ex}

Let $\Gamma$ be the graph constructed as in Figure \ref{fig:graph_example} using the multivector fields $\xi_1,\xi_2,\xi_3\in \Gamma(\bigwedge^\bullet TM)$ with $\vert\xi_1\vert=5,\vert\xi_2\vert=4$, and $\vert\xi_3\vert=2$. Then we get 
\begin{multline}
    \calV_\Gamma(\xi_1v^{\ell_1}\xi_2v^{\ell_2}\xi_3v^{\ell_3}\mid [a_0\otimes a_1\otimes a_2\otimes a_3\otimes a_4])\\
    = \sum \xi_1^{i_1i_2i_3i_4i_5}\de_{i_1}\xi_2^{j_1j_2j_3j_4}\de_{j_1}\xi_3^{m_1m_2}\de_{i_2}a_0\de_{i_3}a_1\de_{i_4}a_2\de_{j_2}a_3\de_{m_1}a_4\theta_{i_5}\theta_{j_3}\theta_{j_4}\theta_{m_2},
\end{multline}
where we sum over all indices and where we set $\theta_i:=\frac{\de}{\de x^{i}}$ for local coordinates $(x^i)$ on $M$.

\begin{figure}[ht]
\centering
\begin{tikzpicture}
\node[vertex, label=above: $\xi_1$] (xi1) at (0,0) {};
\node[vertex, label=above: $\xi_2$] (xi2) at (2,1) {};
\node[vertex, label=right: $\xi_3$] (xi3) at (4,0) {};
\node[vertex, label=below: $a_0$] (a0) at (-1.5,-1) {};
\node[vertex, label=below: $a_1$] (a1) at (-0.5,-1) {};
\node[vertex, label=below: $a_2$] (a2) at (0.5,-1) {};
\node[vertex, label=below: $a_3$] (a3) at (2,-1) {};
\node[vertex, label=below: $a_4$] (a4) at (3,-1) {};
\node[circle,draw=black, fill=white, inner sep=0pt,minimum size=5pt] (w1) at (-1,1.5) {};
\node[circle,draw=black, fill=white, inner sep=0pt,minimum size=5pt] (w2) at (3,2) {};
\node[circle,draw=black, fill=white, inner sep=0pt,minimum size=5pt] (w4) at (5,-1) {};
\node[circle,draw=black, fill=white, inner sep=0pt,minimum size=5pt] (w3) at (4.5,1.5) {};
\draw[fermion] (xi1) -- (a0);
\draw[fermion] (xi1) -- (a1);
\draw[fermion] (xi1) -- (a2);
\draw[fermion] (xi1) -- (xi2);
\draw[fermion] (xi1) -- (w1);
\draw[fermion] (xi2) -- (a3);
\draw[fermion] (xi2) -- (xi3);
\draw[fermion] (xi2) -- (w2);
\draw[fermion] (xi2) -- (w3);
\draw[fermion] (xi3) -- (w4);
\draw[fermion] (xi3) -- (a4);
\end{tikzpicture}
\caption{Example of a graph $\Gamma$.}
\label{fig:graph_example}
\end{figure}
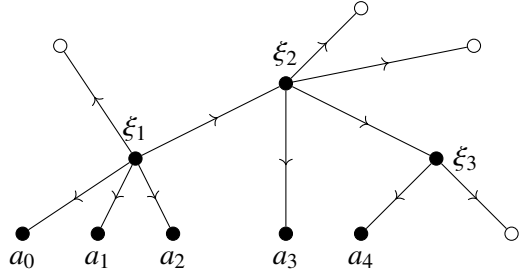

\end{ex}

To compute the configuration integrals, we want to make a degree count, i.e. we want the form degree to be equal to the dimension of the configuration space. Let $\Sigma$ be a manifold with boundary and define the \emph{\textbf{configuration space}} of $n$ points in the bulk and $m$ points on the boundary by
\begin{equation}
    \label{eq:conf_space}
    \mathsf{Conf}_{n,m}(\Sigma):=\{(x_1,\ldots,x_n,y_1,\ldots,y_m)\in \mathrm{int}(\Sigma^n)\times (\de\Sigma)^n\mid x_i\not=x_j,\, y_i\not=y_j\,\forall i\not=j\}.
\end{equation}
Moreover, denote by $\mathsf{C}_{n,m}(\Sigma)$ the \emph{\textbf{FMAS-compactification}} \cite{FulMacPh,AS2} of $\mathsf{Conf}_{n,m}(\Sigma)$ (or of its quotient with respect to the corresponding group action).
Let now $\Sigma=\mathbb{D}$ and fix the point $1$ on $\de\mathbb{D}$. Then we have to work on the section space 
\begin{equation}
\label{eq:section_space}
\mathsf{C}^0_{n,m}(\mathbb{D}):=\{(z,t)\in (\mathrm{int}(\mathbb{D}))^n\times (\de\mathbb{D})^m\mid z_i\not=z_j\,(i\not=j),\, 0<\arg(t_1)<\dotsm <\arg(t_m)<2\pi\}.
\end{equation}
The space \eqref{eq:section_space} has dimension $2n+m$. Moreover, the number $m$ represents the amount of points on the boundary distinct from the fixed point $1$, i.e. the total amount of points on the boundary is $m+1$. In fact, \eqref{eq:section_space} is equal to the set $\{(z,t)\in \mathsf{C}_{n,m+1}(\mathbb{D})\mid t_0=1\}$ for $m\geq 1$.



As already mentioned, we have an $S^1$-action on the disk. Instead of working with the quotient of the configuration space by $PSL_2(\R)$, we will work with \emph{\textbf{equivariant differential forms}}, which arise from the equivariant BV construction of the Poisson Sigma Model within the Feynman graph expansion.

\subsection{Equivariant differential forms and equivariant Stokes' theorem}
We want to work with equivariant differential forms with respect to the $S^1$-action on the disk. We define them as
\begin{equation}
    \Omega_{S^1}^\bullet(\mathbb{D}):=\Omega^\bullet(\mathbb{D})^{S^1}[u],
\end{equation}
where the differential is given by $\dd_{S^1}:=\dd-u\iota_\mathbf{v}$. Here $\mathbf{v}\in \Gamma(T\mathbb{D})$ denotes the image of the infinitesimal vector field $\frac{\dd}{\dd t}$, which is the generator of the infinitesimal action $\R\frac{\dd}{\dd t}\longrightarrow \Gamma(T\mathbb{D})$. Now consider a differential form $\omega$ on the configuration space $\mathsf{C}^0_{n,m}(\mathbb{D})$. We want to describe the boundary of the configuration space. Let $S$ be a subset of $\bar n\geq 2$ points in the bulk which collapse at a point in the bulk of the disk. Then the \emph{\textbf{stratum of type I}} is given by 
\begin{equation}
    \de_S\mathsf{C}_{n,m}(\mathbb{D})\cong\mathsf{C}_{\bar n}(\mathbb{C})\times \mathsf{C}^0_{n-\bar n+1,m}(\mathbb{D}).
\end{equation}
The \emph{\textbf{stratum of type II}} is constructed as follows. Let $S$ be the subset of $\bar n$ points in the bulk and $T$ the subset of $\bar m$ points on the boundary which collapse at a point on the boundary of the disk. Hence we get the stratum
\begin{equation}
    \de_{S,T}\mathsf{C}_{n,m}(\mathbb{D})\cong \mathsf{C}_{\bar n, \bar m}(\mathbb{H})\times \mathsf{C}^0_{n-\bar n,m-\bar m+1}(\mathbb{D}),
\end{equation}
where $\mathbb{H}$ denotes the upper half plane.
\begin{thm}[Equivariant Stokes \cite{CattaneoFelder2010}]
\label{thm:equiv_Stokes}
Let $\omega\in \Omega^\bullet_{S^1}(\mathsf{C}_{n,m+1}(\mathbb{D}))$. Denote also by $\omega$ its restriction on $\mathsf{C}^0_{n,m}(\mathbb{D})\subset \mathsf{C}_{n,m+1}(\mathbb{D})$. Denote by $\omega^\de$ its restriction to the coboundary $1$ strata $\de_i\mathsf{C}^0_{n,m}(\mathbb{D})$. Then 
\begin{equation}
    \int_{\mathsf{C}^0_{n,m}(\mathbb{D})}\dd_{S^1}\omega=\sum_i\int_{\de_i\mathsf{C}^0_{n,m}(\mathbb{D})}\omega^\de -u \int_{\mathsf{C}_{n,m+1}(\mathbb{D})}\omega
\end{equation}
\end{thm}


\subsection{Weights of graphs}
We will consider a \emph{\textbf{propagator}} $\mathscr{P}$ on $\mathbb{D}\times \mathbb{D}\setminus \mathsf{diag}$, where $\mathsf{diag}:=\{(z,z)\mid z\in\mathbb{D}\}\subset \mathbb{D}\times \mathbb{D}$ denotes the diagonal on the disk. The propagator will be a 1-form on the configuration space of the disk. In particular we have
\begin{equation}
\label{eq:propagator}
    \mathscr{P}(z,w):=\frac{1}{4\pi \I}\left(\dd \log \frac{(z-w)(1-z\bar w)}{(\bar z-\bar w)(1-\bar z w)}+z\dd\bar z-\bar z\dd z\right).
\end{equation}
Note that this propagator is equivariant under the $S^1$-action, hence $\mathscr{P}\in \Omega^1(\mathbb{D})^{S^1}$. 
\begin{rem}
An important fact \cite{BC,CMR2} of the propagator is 
\begin{equation}
\label{eq:edge_split}
    \dd \mathscr{P}(z_1,z_2)=\pm\sum_j \pi_1^*\chi_j\land\pi_2^*\chi^j=\pm \Delta_1(\mathsf{x}_1\land\mathsf{e}_2),
\end{equation}
where $\pi_1,\pi_2$ are the projections to the first and second factor respectively. Here $\chi_j,\chi^j$ are representatives of the cohomology classes and their duals respectively, such that $\int_\mathbb{D}\chi_i\land \chi^j=\delta_i^j$. 
\end{rem}
Computing this directly, we get
\begin{align}
\dd_{S^1}\mathscr{P}&=\dd \mathscr{P}-u\iota_{\textbf{v}}\mathscr{P}\\
&=\frac{1}{4\pi\I}\dd (z\dd\bar z-\bar z\dd z)-u\iota_{\textbf{v}}\mathscr{P}\\
\label{eq:diff_prop_3}
&=-\pi_1^*\left(\frac{\I}{2\pi}\dd z\land \dd \bar z+u(1-\vert z\vert^2)\right).
\end{align}
The first term of \eqref{eq:diff_prop_3} is a volume form on the disk and hence a representative of the cohomology class, hence the whole is a representative of the equivariant cohomology class.

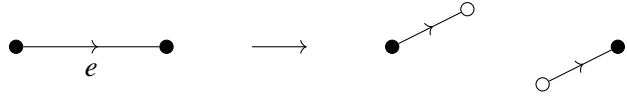
\begin{figure}[ht]
\centering
\begin{tikzpicture}
\node[vertex] (1) at (0,0) {};
\node[vertex] (2) at (2,0) {};
\node[] (a1) at (3,0) {};
\node[] (a2) at (4,0) {};
\node[vertex] (3) at (5,0) {};
\node[circle,draw=black, fill=white, inner sep=0pt,minimum size=5pt] (w1) at (6,0.5) {};
\node[circle,draw=black, fill=white, inner sep=0pt,minimum size=5pt] (w2) at (7,-0.5) {};
\node[vertex] (4) at (8,0) {};
\node[] (e) at (1,-0.3) {$e$};
\draw[fermion] (1) -- (2);
\draw[->] (a1) -- (a2);
\draw[fermion] (3) -- (w1);
\draw[fermion] (w2) -- (4);
\end{tikzpicture}
\caption{Edge split}
\label{fig:edge_split}
\end{figure}

Graphically, this corresponds to the fact that if the de Rham differential acts on an edge of a graph between two (black) vertices (which represents a propagator), it will split into residual fields (see Figure \ref{fig:edge_split}). This can be extended to the equivariant differential $\dd_{S^1}$. The white vertices mentioned in the graph construction before are actually represented by zero modes on $\mathbb{D}$. More precisely, we have the following Lemma.
\begin{lem}[e.g. \cite{CattaneoFelder2010,CMR2}]
\label{lem:edge_split}
Let $\de_e\Gamma$ be the graph which is obtained from the graph $\Gamma$ by adding a white vertex $\circ$ and replacing the edge $e\in E_b(\Gamma)$ connecting two black vertices by an edge originating at the same vertex as $e$ but ending at the white vertex $\circ$. Then
\begin{equation}
    \dd_{S^1}\omega_\Gamma=\sum_{e\in E_b(\Gamma)}(-1)^{\vert E_b(\Gamma)\vert}\omega_{\de_e\Gamma}.
\end{equation}
\end{lem}

The represented zero modes are parametrized by the formal variable $v^\ell$ attached to each vertex. The weight of a graph $\Gamma\in \mathcal{G}_{(k_1,\ldots,k_n),m}$ is then computed by
\begin{equation}
\label{eq:weight}
    w_\Gamma=\frac{1}{k_1!\dotsm k_n!}\int_{\mathsf{C}^0_{n,m}(\mathbb{D})}\omega_\Gamma.
\end{equation}
The \emph{\textbf{equivariant cohomology}} $H^\bullet_{S^1}(\mathbb{D})$ is generated by the constant function $1$. Moreover, the \emph{\textbf{relative equivariant}} cohomology $H^\bullet_{S^1}(\mathbb{D},\de\mathbb{D})$ is generated by the class of 
\begin{equation}
\label{eq:zero_modes_prop}
    \phi(z,u):=\frac{\I}{2\pi}\dd z\land \dd\bar z+u(1-\vert z\vert^2).
\end{equation}
\begin{rem}
Note that with this notation we have $\dd_{S^1}\mathscr{P}=-\pi_1^*\phi$.
\end{rem}

The differential form $\omega_\Gamma\in \Omega^{2n+m}_{S^1}(\mathsf{C}^0_{n,m}(\mathbb{D}))$ is given by
\begin{equation}
    \omega_\Gamma=\bigwedge_{i\in V_1(\Gamma)}\bigwedge_{(i,j)\in E_b(\Gamma)}\mathscr{P}(z_i,z_j)\bigwedge_{i\in V_1(\Gamma)}\phi(z_i,u)^{r_i},
\end{equation}
where the number $r_i$ is given by the degree of the vertex $i$ plus the amount of white vertices attached to it. 
Moreover, we have the following lemma.

\begin{lem}[\cite{K,CattaneoFelder2010}]
\label{lem:Kontsevich}
For all $z,z'\in \mathbb{D}$ we have 
\begin{equation}
\label{eq:vanishing1}
    \int_{w\in \mathbb{D}}\mathscr{P}(z,w)\land\mathscr{P}(w,z')=0.
\end{equation}
Moreover, for all $z\in \mathbb{D}$ we have 
\begin{equation}
\label{eq:vanishing2}
    \int_{w\in \mathbb{D}}\mathscr{P}(z,w)\land\phi(w,u)=0.
\end{equation}
\end{lem}
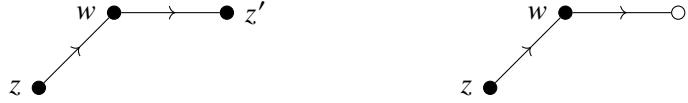
\begin{figure}[ht]
\centering
\begin{tikzpicture}
\node[vertex,label=left: $z$] (z1) at (0,0) {};
\node[vertex,label=left: $w$] (w1) at (1,1) {};
\node[vertex,label=right: $z'$] (z') at (2.5,1) {};
\draw[fermion] (z1) -- (w1);
\draw[fermion] (w1) -- (z');
\node[vertex,label=left: $z$] (z2) at (6,0) {};
\node[vertex,label=left: $w$] (w2) at (7,1) {};
\node[circle,draw=black, fill=white, inner sep=0pt,minimum size=5pt] (z'2) at (8.5,1) {};
\draw[fermion] (z2) -- (w2);
\draw[fermion] (w2) -- (z'2);
\end{tikzpicture}
\caption{The first picture corresponds to the integrand of \eqref{eq:vanishing1} and the second picture corresponds to the one of \eqref{eq:vanishing2}. Graphs with such a vertex $w$ vanish.}
\label{fig:vertex_one_arrow_in_one_arrow_out}
\end{figure}

\section{Main results}
\label{sec:main_results}
\subsection{Proof of the trace property}

\begin{thm}
\label{thm:is_a_trace}
The map \eqref{eq:globalized_trace} is a trace on the algebra $(C^\infty_c(M)[\![ \hbar ]\!],\star)$.
\end{thm}

\begin{proof}
This follows by the fact that 
\begin{equation}
\calV^{\pi^\varphi_\hbar}\colon (CC^-_{-\bullet}(A_\hbar),b+uB)\longrightarrow (\calT^{-\bullet}_{poly}(M)[u][\![\hbar]\!],u\Div_{\Omega})
\end{equation}
is a chain map, which follows from Theorem \ref{thm:equiv_Stokes} and Lemma \ref{lem:edge_split}. Using the construction with the Poisson Sigma Model, the trace property follows from \eqref{eq:path_int_kontsevich} and the constructions in Section \ref{subsec:construction_via_PSM}.

Indeed, consider the observable $O_{\rho(\mathsf{T}\varphi^*f),\rho(\mathsf{T}\varphi^*g)}$ for $f,g\in C_c^\infty(M)[\![ \hbar ]\!]$. Note that the configuration integrals are considered on the section space where the point $1$ is fixed on the boundary, labeled by the observable $O_{\rho(\mathsf{T}\varphi^*f)}$. We consider another point $0$ on the boundary, which is not fixed, labeled by the observable $O_{\rho(\mathsf{T}\varphi^*g)}$. Moreover, we have some additional $m-1$ boundary points labeled by $\gamma$. Note that there are boundary strata of the configuration space where $g$ collides to $f$ from the left and one where it collides from the right. Recall that the dimension of the configuration space $\mathsf{C}^0_{n,m}(\mathbb{D})\subset \mathsf{C}_{n,m+1}(\mathbb{D})$ is given by $2n+m$. Without the point $0$ we would have that the dimension is equal to $2n+m-1$, which has to be the same as the form degree of the differential form $\omega_\Gamma$ within the configuration integral for any graph $\Gamma\in \calG_{(k_1,\ldots,k_n),m}$. Hence, we look at its equivariant differential $\dd_{S^1}\omega_\Gamma$ and apply the equivariant Stokes' theorem (Theorem \ref{thm:equiv_Stokes}). Using \eqref{eq:graph_repr} and \eqref{eq:weight}, we can write 
\begin{multline}
    \calV_n^{\pi^\varphi_\hbar}(R\dotsm R\mid \rho(\mathsf{T}\varphi^*f))=\sum_{\Gamma\in \calG_{\mathbf{k},m}}\left(\int_{\mathsf{C}^0_\Gamma(\mathbb{D})}\dd_{S^1}\omega_\Gamma\right)\calV_\Gamma^{\pi^\varphi_\hbar}(R\dotsm R\mid \rho(\mathsf{T}\varphi^*f))\\
    =\sum_{\Gamma\in\calG_{\mathbf{k},m}}\left(\sum_{\Gamma'<\Gamma}\int_{\mathsf{C}_{\Gamma'<\Gamma}(\mathbb{H})\times \mathsf{C}^0_{\Gamma\setminus \Gamma'}(\mathbb{D})}\omega^\de_{\Gamma'}-u\int_{\mathsf{C}_{\Tilde{\Gamma}}(\mathbb{D})}\omega_{\Tilde{\Gamma}}\right)\calV_\Gamma^{\pi^\varphi_\hbar}(R\dotsm R\mid \rho(\mathsf{T}\varphi^*f)),
\end{multline}
where $\mathbf{k}=(k_1,\ldots,k_n)$ and $\Gamma'<\Gamma$ is a subgraph of $\Gamma$, where $n'<n$ points collapse in the bulk and $m'<m$ collapse on the boundary. Moreover, $\Tilde{\Gamma}$ is a graph whose vertex set satisfies $\vert V(\Tilde{\Gamma})\vert=\vert V(\Gamma)\vert+1$ with the same amount of vertices in the bulk and on the boundary plus an additional vertex on the boundary.
Note that by setting $u=0$, Theorem \ref{thm:equiv_Stokes} reduces to the usual Stokes' theorem for corners.
The dimension of the configuration space $\mathsf{C}_{n,m}(\mathbb{H})$ modulo scaling and translation is given by $2n+m-2$. This has to be equal to the form degree of the differential form we want to integrate. Let $p$ be the amount of vertices labeled by $\pi_\hbar^\varphi$ and $r$ the amount of vertices labeled by $R$. Then we have 
\begin{align}
    \begin{split}
        2n+m-2&=2p+r,\\
        n&=p+r.
    \end{split}
\end{align}
This implies three different cases (see Figure \ref{fig:three_cases});
\begin{itemize}
    \item $r=2,m=0$,
    \item $r=m=1$,
    \item $r=0,m=2$.
\end{itemize}

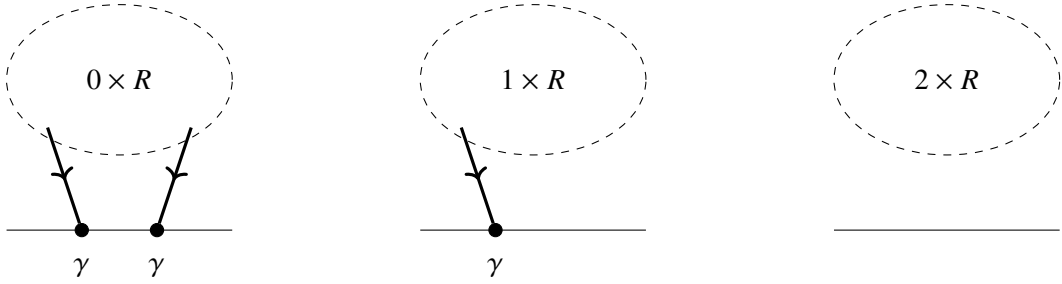
\begin{figure}[ht]
\centering
\begin{tikzpicture}[scale=0.5]
\node[] (a1) at (-9,-3){$\gamma$};
\node[] (a2) at (-7,-3){$\gamma$};
\node[] (R1) at (-8,2){$0\times R$};
\node[] (x1) at (-10,1){};
\node[] (x2) at (-6,1){};
\node[vertex] (y1) at (-9,-2){};
\node[vertex] (y2) at (-7,-2){};
\draw (-11,-2) -- (-5,-2);
\draw[dashed] (-8,2) ellipse (3cm and 2cm);
\draw[very thick, fermion] (x1) -- (y1);
\draw[very thick, fermion] (x2) -- (y2);
\node[] (a) at (2,-3){$\gamma$};
\node[] (R2) at (3,2){$1\times R$};
\node[] (x3) at (1,1){};
\node[] (x4) at (5,1){};
\node[vertex] (y3) at (2,-2){};
\draw (0,-2) -- (6,-2);
\draw[dashed] (3,2) ellipse (3cm and 2cm);
\draw[very thick,fermion] (x3) -- (y3);
\node[] (R3) at (14,2){$2\times R$};
\draw (11,-2) -- (17,-2);
\draw[dashed] (14,2) ellipse (3cm and 2cm);
\end{tikzpicture}
\caption{Illustration of the three different cases. The dashed ellipses represent a graph $\Gamma$ in the bulk of the disk with either $r=0$ (first picture), $r=1$ (second picture), or $r=2$ (third picture). Note that in each picture $p$ can be arbitrary. The thick arrows denote the fact that there can be arbitrarily many incoming arrows, depending on the combinatorics.}
\label{fig:three_cases}
\end{figure}
Summing over all these graphs, the third picture will exactly correspond to $F=F(R,R)$, the curvature of the deformed Grothendieck connection $\calD$, the second picture to $\calD \gamma$ and the first picture is exactly the star product $\gamma\star \gamma$. Thus, summing them together we get a contribution
\begin{equation}
    F+\calD\gamma+\gamma\star\gamma=0,
\end{equation}
and hence these terms vanish. Hence the only strata that survive within the boundary of the configuration space are the ones where $g$ approaches $f$ from the left and from the right, so by \cite{CF1} we get the boundary contribution $g\star f-f\star g$.

In fact, for any $\xi=\xi_1\dotsm \xi_n\in (\Sym^n\calT^{\bullet+1}_{poly}(M)[u],\delta_\Omega)$ and $a\in CC^-_m(A)$, we have 
\begin{multline}
\label{eq:chain_map}
    \calV_n(\delta_\Omega\xi\mid a)+(-1)^{\vert\xi\vert+m}\calV_n(\xi\mid (b+uB)a)\\
    +\sum_{k=0}^{n-1}\sum_{\sigma\in S_{k,n-k}}(-1)^{\vert\xi\vert-1}\varepsilon(\sigma,\xi)\calV_k(\xi_{\sigma(1)}\dotsm \xi_{\sigma(k)}\mid \calU_{n-k}(\bar\xi_{\sigma(k+1)}\dotsm \bar\xi_{\sigma(n)}))\cdot a)\\
    +\sum_{i<j}\varepsilon_{ij}\calV_{n-1}((-1)^{\vert\xi_i\vert-1}[\xi_i,\xi_j]_{SN}\cdot \xi_1\dotsm \bar\xi_i\dotsm \bar\xi_j\dotsm \xi_n\mid a)=\Div_\Omega\calV_n(\xi\mid a),
\end{multline}
where $\bar\xi_i$ denotes the projection of $\xi_i$ to $\calT^{\bullet+1}_{poly}(M)$, $S_{p,q}\subset S_{p+q}$ is the set of $(p,q)$-shuffles and the signs $\varepsilon(\sigma,\xi),\varepsilon_{ij}$ are the Koszul signs coming from the permutation of the $\xi_i$, and $\vert\xi\vert=\sum_{i}\vert\xi_i\vert$. Note that $\delta_\Omega$ is extended to a degree $+1$ derivation on $\Sym\calT^{\bullet+1}_{poly}(M)[u]$. The maps $\calU_k\colon \Sym^k\calT^{\bullet+1}_{poly}(M)\longrightarrow \calD^{\bullet+1}_{poly}(M)$ are the Taylor components of Kontsevich's $L_\infty$-morphism.

Indeed, one can show that for any $a=[a_0\otimes\dotsm\otimes a_m]\in C_{-m}(A)$, $\Gamma\in\calG_{\mathbf{k},m}$ and $\xi=\xi_1\dotsm \xi_n$ with $\xi_i\in \Gamma(\bigwedge^{k_i}TM)$ we have\footnote{We use the same notation as in Lemma \ref{lem:edge_split}.}
\begin{equation}
    \Div_\Omega\calV_n(\xi\mid a)-\calV_n(\delta_\Omega\xi\mid a)=\sum_{(\Gamma,e)}(-1)^{\vert E_b(\Gamma)\vert}w_{\de_e\Gamma}\calV_\Gamma(\xi\mid a),
\end{equation}
by identifying $\Gamma(\bigwedge^n TM)$ with $C^\infty(M)[\theta_1,\ldots,\theta_n]$, where $\theta_i$ are odd variables such that $\Div_\Omega=\sum_{1\leq i\leq n}\frac{\de^2}{\de t_i\de\theta_i}$. In fact, we have 
\begin{equation}
\label{eq:rel1}
    \sum_{e\in E_b(\Gamma)}(-1)^{\vert E_b(\Gamma)\vert}w_{\de_e\Gamma}=\sum_{i}\int_{\de_i\mathsf{C}^0_{n,m}(\mathbb{D})}\omega^\de_\Gamma-u\sum_{k=0}^m(-1)^{km}\int_{\mathsf{C}^0_{n,m+1}(\mathbb{D})}j^*_k\omega_\Gamma,
\end{equation}
where $j_k$ is defined as follows: Define a map 
\begin{align}
\begin{split}
    j_0\colon \mathsf{C}^0_{n,m}(\mathbb{D})&\longrightarrow \mathsf{C}_{n,m}(\mathbb{D})\\
    (z,1,t_1,\ldots,t_m)&\longmapsto (z,t_1,\ldots,t_m)
\end{split}
\end{align}
Moreover, define a map 
\begin{align}
\begin{split}
    \lambda\colon \mathsf{C}^0_{n,m}(\mathbb{D})&\longrightarrow \mathsf{C}^0_{n,m}(\mathbb{D})\\
    (z_1,\ldots,z_n,1,t_1,\ldots,t_m)&\longmapsto (z_1,\ldots,z_n,1,t_m,t_1,\ldots,t_{m-1})
\end{split}
\end{align}
Then the collection $j_k:= j_0\circ \underbrace{\lambda\circ\dotsm\circ\lambda}_{k \text{ times}}$, for $k=0,1,\ldots,m-1$ defines an embedding 
\begin{equation}
j\colon \mathsf{C}^0_{n,m}(\mathbb{D})\sqcup\dotsm \sqcup\mathsf{C}^0_{n,m}(\mathbb{D})\longhookrightarrow \mathsf{C}_{n,m}(\mathbb{D}).
\end{equation}
Moreover, note that
\begin{equation}
    \int_{\mathsf{C}_{n,m+1}(\mathbb{D})}\omega=\sum_{k=0}^m(-1)^{km}\int_{\mathsf{C}^0_{n,m+1}(\mathbb{D})}j^*_k\omega,
\end{equation}
and the second term on the right-hand side of \eqref{eq:rel1} is given by $\calV_{n+1}(\xi\mid Ba)$.
Let us look at the boundary integral in the first term of the right hand side of \eqref{eq:rel1}. As argued in \cite{CattaneoFelder2010}, one can show that treating the boundary strata of type I, the only remaining term will be the sum in \eqref{eq:chain_map} containing the Schouten--Nijenhuis bracket. The strata of type II will give a contribution as the sum in \eqref{eq:chain_map} containing Kontsevich's $L_\infty$-morphism and a term $\calV_{n-1}(\xi\mid ba)$. 

Note that \eqref{eq:rel1} together with \eqref{eq:trace_condition_1}, \eqref{eq:edge_split} and Lemma \ref{lem:edge_split} ensure that $\tr(f\star g)=\tr(g\star f)$ since $\Div_\Omega\pi^\varphi_\hbar=0$.

Using Equation \eqref{eq:chain_map}, we get that the twist of $\calV$ by $\pi^\varphi_\hbar$ is indeed a chain map. Recall from Section \ref{sec:traces_and_algebraic_index} that the zeroth cyclic homology $CH_0(A_\hbar)$ is isomorphic to the zeroth Hochschild homology $HH_0(A_\hbar)$, which is again isomorphic to the zeroth Poisson homology $HP_0(M)$. Hence the chain map $\calV^{\pi^\varphi_\hbar}$ induces a map 
\begin{equation}
    C^\infty(M)[\![\hbar]\!][u,u^{-1}]\cong CH_0(A_\hbar)\cong HH_0(A_\hbar)\xrightarrow{\calV^{\pi^\varphi_\hbar}} HP_0(M)\longrightarrow \R(\!(\hbar)\!)[u,u^{-1}],
\end{equation}
given by integration as in \eqref{eq:Poisson_homology_map}. 
\end{proof}

\subsection{Relation to the Tamarkin--Tsygan theorem}
\label{subsec:tamrakin-tsygan}

\begin{thm}
The trace formula in \eqref{eq:globalized_trace} evaluated at any periodic cyclic chain $c\in PC_{-m}(A_\hbar)$ is given by \eqref{eq:Tamarkin-Tsygan}.
\end{thm}

\begin{proof}
Note that since the Lie derivative with respect to the Poisson tensor $\pi$ is defined by $L_\pi:=\dd\circ\iota_\pi-\iota_\pi\circ\dd$, we get an isomorphism of complexes
\begin{align}
\begin{split}
    \label{eq:iso_complex}
    (\Omega^{-\bullet}(M,\R)[\![ \hbar ]\!][u,u^{-1}],L_\pi+u\dd)&\longrightarrow (\Omega^{-\bullet}(M,\R)[\![ \hbar ]\!][u.u^{-1}],u\dd)\\
    \alpha&\longmapsto \alpha\exp\left(\iota_\pi/u\right).
\end{split}
\end{align}
Let $A_t$ be a 1-parameter family of algebras given by $A[\![t]\!]$ as an $\R[t]$-module.
Denote by $\calV^\pi:=\sum_{n=0}^\infty\frac{1}{n!}\calV_n^\pi$.
Consider the \emph{\textbf{Gauss--Manin connection}} on the periodic cyclic cohomology viewed as a vector bundle over the parameter space (see e.g \cite{Getzler1993,Tsygan2007,CattaneoFelderWillwacher2011}). In \cite{CattaneoFelderWillwacher2011} it was shown that for a 1-parameter family $\pi_t$ of solutions to the Maurer--Cartan equation, e.g. with polynomial dependence $\pi_t=t\pi$ for a Poisson tensor $\pi$, and a cyclic cycle $c_t\in PC_{-m}(A_t)$, which is horizontal with respect to the Gauss--Manin connection, the class of $((\exp\left(\iota_{\pi_t}/u\right)\calV^{\pi_t})(c_t)$ in $\bigoplus_{j\geq 0}H^{m+2j}(M,\R)[\![ \hbar ]\!]u^j$ is independent of $t$. Denote by $\Tilde{\calV}^{\pi}:=\exp\left(\iota_{\pi_t}/u\right)\calV^{\pi_t}$ the image of $\calV^{\pi_t}$ under the isomorphism \eqref{eq:iso_complex} with formal Poisson structure $\pi_t$. Since $[\Tilde{\calV}^{\pi_t}(c_t)]$ is independent of $t$, we can set the Poisson tensor to be zero. In our case we have $t=\hbar$ and this will allow us to put $\hbar=0$ in $\pi^\varphi_\hbar$ and to merge all the $\pi$ vertices at zero (see Figure \ref{fig:wheel_graph_general}). This will produce wheel graphs as considered e.g. in \cite{VanBergh2009,Willwacher2007}. Let us denote the weight for a wheel graph with $j$ vertices by $w_j$.

Note that the curvature of the Grothendieck connection is contained in the $R$-vertices (see Section \ref{subsec:Grothendieck_conn_symplectic_manifolds}). Hence, considering the Propagator $\mathscr{P}$ on the disk we can compute the weight of a wheel diagram with $j$ black vertices. We will get 
\begin{equation}
    w_{j}=\int_{\mathsf{C}^0_{j+1,0}(\mathbb{D})}\mathscr{P}(z_1,z_2)\land\dotsm \land \mathscr{P}(z_j,z_1)\land\bigwedge_{k=1}^{j}\mathscr{P}(0,z_k)
\end{equation}
where $z_1,\ldots,z_j$ are the vertices labeled by $R$.
Moreover, if we recall that $R_\ell(x,y)=R^{k}_\ell(x,y)\frac{\de}{\de y^k}$ and $R^k(x,y):=R^k_\ell(x,y) \dd x^\ell$, we get a differential form  
\begin{equation}
    \calV_j(R\dotsm R\mid 1)=\sum\de_{\ell_1}\de_{k_j}R^{k_1}_{\ell_1}(x,y)\dd x^{\ell_1}\land\de_{\ell_2}\de_{k_1}R^{k_2}_{\ell_2}(x,y)\dd x^{\ell_2}\land\dotsm\land \de_{\ell_j}\de_{k_{j-1}}R^{k_j}_{\ell_j}(x,y)\dd x^{\ell_j},
\end{equation}
where we sum over all indices. Thus permuting everything into the right place we get $\tr(R^j)$, where by abuse of notation we also denote by $R$ the appearing curvature. The permutation will give a sign
\begin{equation}
\prod_{k=1}^{j-1}(-1)^s=(-1)^{\sum_{s=1}^{j-1}s}=(-1)^{j(j-1)/2}.
\end{equation}

To get the correct form degree and be consistent with the isomorphism \ref{eq:iso_complex}, we will need a factor of $u^j$. Indeed, note that $\phi(0,u)=u$, which in fact appears for any $\pi_\hbar^\varphi$-vertex and hence merging this $j$ times we get a factor $u^j$.

One can easily see that the propagator $\mathscr{P}$ will reduce to Kontsevich's angle propagator $\mathscr{P}(0,z_i)=\frac{1}{2\pi}\dd\arg(z_i)$ and hence $w_j$ vanishes if $j$ is odd. 
Note that if $j$ is even, we have
\begin{equation}
    \int_\mathbb{D}\bigwedge_{k=1}^j\mathscr{P}(0,z_k)=\frac{1}{(2\pi)^j}\int_{\mathbb{D}}\bigwedge_{k=1}^j\dd\arg(z_k)=\frac{1}{(2\pi)^j}\prod_{k=1}^j\int_{\mathbb{D}}\dd \arg(z_k)=\prod_{k=1}^j\frac{1}{k}=\frac{1}{j!}.
\end{equation}
Therefore, as it was computed in \cite{VanBergh2009,Willwacher2007}, we get 
\begin{equation}
    w_j=\int_{\mathsf{C}^0_{j,0}(\mathbb{D})}\mathscr{P}(z_1,z_2)\land\dotsm \land \mathscr{P}(z_{j},z_1)\land\bigwedge_{k=1}^j\mathscr{P}(0,z_k)=-(-1)^{j(j-1)/2}\frac{B_j}{(2j)j!},
\end{equation}
where $B_j$ are the \emph{\textbf{Bernoulli numbers}}\footnote{Note that this makes indeed sense since the $B_j$ is zero for odd $j\not=1$. The case of $j=1$ is not relevant since $j$ has to be at least two.}. Hence we have
\begin{equation}
    [\Tilde{\calV}^{\pi=0}(1)]=\left[\exp\left(\sum_{j\geq 2}w_ju^j\calV_j(R\dotsm R\mid 1)\right)\right].
\end{equation}
Thus we get 
\begin{equation}
    [\Tilde{\calV}^{\pi=0}(1)]=\left[\exp\left(-\sum_{j\geq 1}\frac{B_{2j}}{(4j)(2j)!}u^{2j}\tr(R^{2j})\right)\right]=\mathrm{det}^{1/2}\left(\frac{uR/2}{\sinh(uR/2)}\right)=\Hat{A}_u(TM).
\end{equation}

Recall from Section \ref{subsec:Kontsevich_Tsygan_formality} that the \emph{\textbf{Connes isomorphism}} $\mathrm{Co}$ is given by
\begin{align}
\begin{split}
\mathrm{Co}\colon PH_{\bullet}(A)&\longrightarrow H^\bullet(M,\R)[u,u^{-1}]\\
[a_0\otimes\dotsm\otimes a_m]&\longmapsto \frac{1}{m!}a_0\dd a_1\land\dotsm \land \dd a_m.
\end{split}
\end{align}

Then for any $c\in PC_{\bullet}(A)$ we can see that one gets
\begin{equation}
    [\Tilde{\calV}^{\pi=0}(c)]=\Hat{A}_u(TM)\mathrm{Co}(c),
\end{equation}
In fact, $\mathrm{Co}(c_0)=\mathrm{Ch}(c)$ is the Chern character of the cyclic cycle $c$. Note that if $c=1$ we get $\mathrm{Co}(1)=1$.
The map $I$ is given by integration $\int_M[\Tilde{\calV}^{\pi}(c)]\exp(\iota_\pi/u)\Omega$ for a chosen volume form $\Omega$ on $M$ and thus 
\begin{equation}
I(c)=\int_M\Hat{A}_u(TM)\mathrm{Ch}(c)\exp\left(\iota_\pi/u\right)\Omega.
\end{equation}

\begin{figure}[ht]
\begin{tikzpicture}[scale=0.8]
\tikzset{Bullet/.style={fill=black,draw,color=#1,circle,minimum size=0.5pt,scale=0.5}}
\node[vertex] (1) at (0,0) {};
\node[vertex] (2) at (1,1) {};
\node[vertex] (3) at (3,1) {};
\node[vertex] (4) at (4,0) {};
\node[vertex] (5) at (4,-1.5) {};
\node[vertex] (6) at (3,-2.5) {};
\node[vertex] (7) at (1,-2.5) {};
\node[vertex] (8) at (0,-1.5) {};
\node[vertex] (center) at (2,-0.75) {};
\node[vertex, label=below: $a_1$] (a1) at (-1.1,-4.7) {};
\node[vertex, label=below: $a_2$] (a2) at (2,-5.75) {};
\node[vertex, label=below: $a_3$] (a3) at (4.9,-4.8) {};
\node[vertex, label=right: $a_4$] (a4) at (6.3,-3.3) {};
\node[Bullet=gray] (pi5) at (4.6,-3.3) {};
\node[Bullet=gray] (pi4) at (6,-2) {};
\node[Bullet=gray] (pi3) at (3,-4.5) {};
\node[Bullet=gray] (pi2) at (1,-4.5) {};
\node[Bullet=gray] (pi1) at (-1,-3) {};
\draw[fermion] (1) -- (2);
\draw[fermion] (2) -- (3);
\draw[fermion] (3) -- (4);
\draw[fermion] (4) -- (5);
\draw[fermion] (5) -- (6);
\draw[fermion] (6) -- (7);
\draw[fermion] (7) -- (8);
\draw[fermion] (8) -- (1);
\draw[fermion2] (pi1) -- (8);
\draw[fermion2] (pi1) -- (a1);
\draw[fermion2] (pi2) -- (a1);
\draw[fermion2] (pi2) -- (a2);
\draw[fermion2] (pi3) -- (a2);
\draw[fermion2] (pi3) -- (a3);
\draw[fermion2] (pi5) -- (a3);
\draw[fermion2] (pi5) -- (a4);
\draw[fermion2] (pi4) -- (a4);
\draw[fermion] (center) -- (1);
\draw[fermion] (center) -- (2);
\draw[fermion] (center) -- (3);
\draw[fermion] (center) -- (4);
\draw[fermion] (center) -- (5);
\draw[fermion] (center) -- (6);
\draw[fermion] (center) -- (7);
\draw[fermion] (center) -- (8);
\draw[fermion] (center) -- (a1);
\draw[fermion] (center) -- (a2);
\draw[fermion] (center) -- (a3);
\draw[fermion] (center) -- (a4);
\draw (2,-0.75) circle [radius=5cm];
\node[vertex,label=right: $1$] (b) at (7,-0.75) {};
\draw[fermion2] (pi4) -- (b);
\end{tikzpicture}
\caption{Illustration of the merging of all $\pi_\hbar^\varphi$-vertices (illustrated in gray) to the center of the disk. The black vertices on the wheel contain the curvature coming from the Grothendieck connection by the 1-forms $R$.}
\label{fig:wheel_graph_general}
\end{figure}
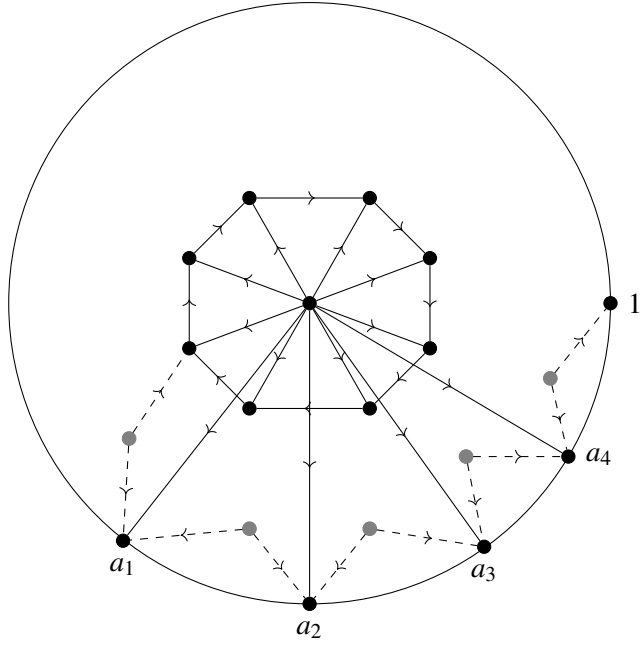

\end{proof}

\subsection{The symplectic case}
In \cite{GLL} a similar construction was considered for symplectic manifolds. They formulate a global trace map using the 1-dimensional Chern--Simons theory\footnote{This is simply the case of topological quantum mechanics with action $S=\int_{S^1} p\dd q$.} within the setting of the BV formalism, by considering solutions of the Quantum Master Equation, and solutions of Fedosov's equation \eqref{eq:fedosov}. 
Moreover, they extend this map to an equivariant one with respect to the $S^1$-action.
Let us give some more details for this construction. Let $(M,\omega)$ be a symplectic manifold of dimension $2d$ and consider a symplectic connection $\nabla$ on $TM$. Let
\begin{equation}
    \Hat{\Omega}^{-\bullet}(TM):= \Hat{\Sym}(T^*M)\otimes \bigwedge^{-\bullet}T^*M,\qquad \bigwedge^{-\bullet}T^*M:=\bigoplus_k\bigwedge^k(T^*M)[k]
\end{equation}
Moreover, define a map 
\begin{align}
\label{eq:Ber_int}
\begin{split}
    \int_{Ber}\colon \Gamma(\Hat{\Omega}^{-\bullet}(TM))&\longrightarrow C^\infty(M)\\
    a(x,y,\hbar)&\longmapsto \int_{Ber}a(x,y,\hbar):=\frac{1}{d!}\left(\frac{1}{2}\omega^{ij}\de_{y^i}\land\de_{y^j}\right)^da(x,y,\hbar)\Big|_{x=y=0},
\end{split}
\end{align}
where $\omega^{ij}$ are the components of $\omega^{-1}$. Let $\Omega^\bullet(M,\Hat{\Omega}^{-\bullet}(TM))$ denote the complex of differential forms with values in $\Hat{\Omega}^{-\bullet}(TM)$. The symplectic connection can be extended to a map 
\begin{equation}
    \nabla\colon \Omega^k(M,\Hat{\Omega}^{-\bullet}(TM))\longrightarrow \Omega^{k+1}(\Hat{\Omega}^{-\bullet}(TM)).
\end{equation}
A degree zero element $S\in\Omega^\bullet(M,\Hat{\Omega}^{-\bullet}(TM))[\![ \hbar ]\!]$ is said to satisfy the Quantum Master Equation if
\begin{equation}
\label{eq:QME}
    \left(\nabla+\I\hbar\Delta+\frac{\I}{\hbar}\dd_{TM}F\right)\exp\left(\frac{\I}{\hbar}S\right)=0, 
\end{equation}
where $\dd_{TM}$ is the de Rham differential on $TM$, $\Delta:=L_\pi=[\dd_{TM},\iota_\pi]$ with $\pi$ the Poisson structure induced by $\omega$ (here $L$ denotes the Lie derivative), and $F$ is the Weyl curvature tensor given as in \eqref{eq:Weyl_curvature}. 
In fact $(\Hat{\Omega}^{-\bullet}(TM),\Delta)$ is a BV algebra like as in Appendix \ref{app_subsec:BV_algebras}, which is why in \cite{GLL} they call $\Hat{\Omega}^{-\bullet}(TM)$ the \emph{\textbf{BV bundle}}.
One can show that if \eqref{eq:QME} is satisfied, the operator $\nabla+\I\hbar+\{S,\enspace\}_\Delta$ is a differential on $\Omega^\bullet(M,\Hat{\Omega}^{-\bullet}(TM))$. Here $\{\enspace,\enspace\}_\Delta$ denotes the odd Poisson bracket defined by $\Delta$.
For a solution $S$ of \eqref{eq:QME}, we define the \emph{\textbf{twisted integration map}} 
\begin{align}
\label{eq:twisted_int}
\begin{split}
    \int_S\colon \Omega^\bullet(M,\Hat{\Omega}^{-\bullet}(TM))&\longmapsto \Omega^\bullet(M,\R)(\!( \hbar )\!)\\
    a&\longmapsto \int_Sa:=\int_{Ber}\exp\left(\frac{\I}{\hbar}S\right)a.
\end{split}
\end{align}
In fact, one can show that 
\begin{equation}
    \int_S\colon (\Omega^\bullet(M,\Hat{\Omega}^{-\bullet}(TM))[\![ \hbar ]\!],\nabla+\I\hbar\Delta+\{S,\enspace\}_\Delta)\longrightarrow (\Omega^\bullet(M,\R)(\!( \hbar )\!),\dd)
\end{equation}
is a cochain map and hence, by composition, we have a map 
\begin{equation}
    \int_M\int_S\colon H^0(\Omega^\bullet(M,\Hat{\Omega}^{-\bullet}(TM))[\![ \hbar ]\!],\nabla+\I\hbar\Delta+\{S,\enspace\}_\Delta)\longrightarrow \R(\!( \hbar )\!).
\end{equation}
Fix a solution $\gamma$ of \eqref{eq:fedosov}. Then one can construct a nilpotent\footnote{i.e. there is some $N\gg0$ such that $\gamma_\infty^N=0$, which is in fact true since the exponential map will terminate for some power.} solution $\gamma_\infty$ of \eqref{eq:QME} as an effective action
\begin{equation}
    \gamma_\infty:=-\I\hbar\log\sum_{\Gamma\in \calG^0}\frac{\hbar^{\ell(\Gamma)}}{\vert\mathrm{Aut}(\Gamma)\vert}\int_{\mathsf{C}_\Gamma(S^1)}\omega_\Gamma(\gamma,\mathscr{P}_{S^1}), 
\end{equation}
where $\calG^0$ denotes the set of all connected graphs, $\ell(\Gamma)$ denotes the number of loops of $\Gamma$, $\mathrm{Aut}(\Gamma)$ denotes the automorphism group of $\Gamma$, and $\omega_\Gamma(\gamma,\mathscr{P}_{S^1})$ a differential form depending on a chosen propagator $\mathscr{P}_{S^1}$ on $S^1$ and $\gamma$. 

Define a map 
\begin{equation}
    [\enspace]_\infty\colon \Omega^\bullet(M,\calW)\longrightarrow\Omega^\bullet(M,\Hat{\Omega}^{-\bullet}(TM))[\![ \hbar ]\!]
\end{equation}
which represents a \emph{\textbf{factorization map}} from local observables on the interval to global observables on $S^1$. 
The trace map in this setting is defined by 
\begin{align}
\label{eq:trace_symplectic}
    \begin{split}
        \tr\colon C^\infty(M)[\![ \hbar ]\!]&\longrightarrow \R(\!( \hbar )\!)\\
        f&\longmapsto \tr(f):=\int_M\int_{\gamma_\infty}[\sigma^{-1}(f)]_\infty,
    \end{split}
\end{align}
where $\sigma$ is the symbol map \eqref{eq:symbol_map}.

For the equivariant formulation, extend the map $\sigma$ to the BV bundle
\begin{equation}
    \sigma\colon \Omega^\bullet(M,\Hat{\Omega}^{-\bullet}(TM))\longrightarrow \Omega^\bullet(M,\R),
\end{equation}
by sending $y^i,\dd y^i\mapsto 0$, and define the $S^1$-equivariantly extended complexes
\begin{equation}
    \Omega^\bullet(M,\calW)^{S^1}:=\left(\Omega^\bullet(M,\calW)[u,u^{-1},\dd t],\nabla+\frac{1}{\hbar}[\gamma,\enspace]_\star-u\iota_{\frac{\dd}{\dd t}}\right),
\end{equation}
where $t$ is the coordinate on $S^1$, and
\begin{equation}
    \Omega^\bullet(M,\Hat{\Omega}^{-\bullet}(TM))^{S^1}[\![ \hbar ]\!]:=\left(\Omega^\bullet(M,\Hat{\Omega}^{-\bullet}(TM))[u,u^{-1}],\nabla+\I\hbar\Delta+\{\gamma_\infty,\enspace\}_\Delta+u\dd_{TM}\right).
\end{equation}
Moreover, one can extend the map $[\enspace]_\infty$ to an equivariant version  
\begin{equation}
    [\enspace]_\infty^{S^1}\colon \Omega^\bullet(M,\calW)^{S^1}\longrightarrow \Omega^\bullet(M,\Hat{\Omega}^{-\bullet}(TM))^{S^1}[\![ \hbar ]\!],
\end{equation}
and show that it still remains a cochain map for the equivariant differentials. Furthermore, one also defines an \emph{\textbf{equivariant twisted integration map}} 
\begin{align}
\label{eq:equiv_twisted_int}
\begin{split}
    \int_{\gamma_\infty}^{S^1}\colon \Omega^\bullet(M,\Hat{\Omega}^{-\bullet}(TM))^{S^1}[\![ \hbar ]\!]&\longrightarrow \Omega^\bullet(M,\R)(\!( \hbar )\!)[u,u^{-1}]\\
    a&\longmapsto \sigma\left(u^d\exp\left(\hbar\iota_\pi/u\right)a\exp\left(\frac{\I}{\hbar}\gamma_\infty\right)\right).
\end{split}
\end{align}
\begin{rem}
In fact one can show that \eqref{eq:equiv_twisted_int} extends \eqref{eq:twisted_int} as 
\begin{equation}
    \lim_{u\to 0}\int_{\gamma_\infty}^{S^1}a=\int_{\gamma_\infty}a, \qquad a\in \Omega^\bullet(M,\Hat{\Omega}^{-\bullet}(TM)).
\end{equation}
\end{rem}
Again, one can show that \eqref{eq:equiv_twisted_int} remains a cochain map with respect to the extended complexes, and in particular the composition 
\begin{equation}
    \int_{\gamma_\infty}^{S^1}[\enspace]_\infty^{S^1}\colon \Omega^\bullet(M,\calW)^{S^1}\longrightarrow \Omega^\bullet(M,\R)(\!( \hbar )\!)[u,u^{-1}]
\end{equation}
is a cochain map. The $S^1$-equivariant trace map is then defined by 
\begin{align}
\label{eq:equiv_trace_map}
\begin{split}
    \tr^{S^1}\colon \Omega^\bullet(M,\calW)^{S^1}&\longrightarrow \R(\!( \hbar )\!)[u,u^{-1}]\\
    f&\longmapsto \tr^{S^1}(f)=\int_M\int_{\gamma_\infty}^{S^1}[f]_\infty^{S^1}.
\end{split}
\end{align}
Moreover, the relation to \eqref{eq:trace_symplectic} is 
\begin{equation}
    \tr(f)=\tr^{S^1}(\dd t\sigma^{-1}(f)).
\end{equation}

\subsection{Feynman graphs for cotangent targets}
\label{subsec:Feynman_graphs_cotangent}
Consider the case of the Poisson Sigma Model with target a cotangent bundle $M=T^*N$ for some manifold $N$. Then by Proposition \ref{prop:linear_R} and Lemma \ref{lem:Kontsevich} the graphs will reduce to a certain class of graphs. We have two different bulk vertices. There are vertices labeled by $\pi_\hbar^\varphi$ and vertices labeled by $R$. The $\pi_\hbar^\varphi$-vertices emanate two arrows, representing $\bar q$- and $\bar p$-derivatives as in Section \ref{subsec:lifting_exp_map_cotangent_bundles}, and there are no arrows arriving at them, since the Poisson structure is constant. The $R$-vertices emanate one arrow and there can be an arbitrary amount of arrows representing $\bar q$-derivatives arriving at them, but by Proposition \ref{prop:linear_R} we can only have at most one arrow representing a $\bar p$-derivative arriving. We also consider vertices on the boundary representing solutions $\gamma$ of \eqref{eq:general_fedosov}. For each of them there are no arrows emanating and arbitrarily many arriving. 

\begin{figure}[ht]
\centering
\begin{tikzpicture}
\tikzset{Bullet/.style={fill=black,draw,color=#1,circle,minimum size=0.5pt,scale=0.5}}
\node[Bullet=gray, label=above: $\pi^\varphi_\hbar$] (pi) at (0,0) {};
\node[vertex, label=right: $R$] (R) at (4,0) {};
\node[] (1) at (-1,-1) {};
\node[] (2) at (1,-1) {};
\node[] (3) at (4,-1) {};
\node[] (u1) at (3,1) {};
\node[] (u2) at (3.5,1) {};
\node[] (u3) at (4,1) {};
\node[] (u4) at (4.5,1) {};
\node[] (u5) at (5,1) {};
\node[vertex ,label=below: $\gamma$] (g) at (8,-1) {};
\node[] (g1) at (6,0) {};
\node[] (g2) at (6.5,0) {};
\node[] (g3) at (7.5,0) {};
\node[] (g4) at (8.5,0) {};
\node[] (g5) at (9,0) {};
\node[] (g6) at (10,0) {};
\draw[->,decorate,decoration={snake,amplitude=.4mm,segment length=2mm,post length=1mm}] (pi) -- (1);
\draw[fermion] (pi) -- (2);
\draw[fermion] (R) -- (3);
\draw[->,decorate,decoration={snake,amplitude=.4mm,segment length=2mm,post length=1mm}] (u1) -- (R);
\draw[fermion] (u2) -- (R);
\draw[fermion] (u3) -- (R);
\draw[fermion] (u5) -- (R);
\draw[->,decorate,decoration={snake,amplitude=.4mm,segment length=2mm,post length=1mm}] (g1) -- (g);
\draw[->,decorate,decoration={snake,amplitude=.4mm,segment length=2mm,post length=1mm}] (g2) -- (g);
\draw[->,decorate,decoration={snake,amplitude=.4mm,segment length=2mm,post length=1mm}] (g3) -- (g);
\draw[fermion] (g4) -- (g);
\draw[fermion] (g5) -- (g);
\draw[fermion] (g6) -- (g);
\draw[dotted] (8.7,-0.3) arc (110:60:0.5); 
\draw[dotted] (4,0.8) arc (90:30:0.5); 
\draw[dotted] (7.2,-0.3) arc (120:80:0.5); 
\draw (7,-1) -- (9,-1);
\end{tikzpicture}    
\caption{The interaction vertices appearing in the cotangent case. The straight arrows represent a $\bar q$-derivative and the wavy arrows represent a $\bar p$-derivative. There are no incoming arrows at the $\pi_\hbar^\varphi$-vertices and exactly two emanating arrows. There are arbitrarily many incoming arrows representing the $\bar q$-derviatives for an $R$-vertex, but at most one arrow representing a $\bar p$-derivative and exactly one arrow emanating. For the $\gamma$-vertices we have arbitrarily many incoming $\bar q$- and $\bar p$-derivatives and no emanating arrows.
}
\label{fig:vertices_for_cotangent}
\end{figure}
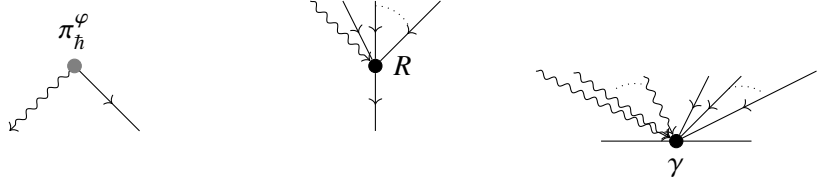

\begin{ex}
Examples of graphs appearing for cotangent targets are given in Figure \ref{fig:ex1} and \ref{fig:ex2}.

\begin{figure}[ht]
\centering
\begin{tikzpicture}
\tikzset{Bullet/.style={fill=black,draw,color=#1,circle,minimum size=0.5pt,scale=0.5}}
\node[vertex, label=left: $R$] (R) at (0,0) {};
\node[Bullet=gray, label=above: $\pi^\varphi_\hbar$] (pi1) at (1.5,2) {};
\node[vertex, label=below: $c$] (f) at (2.5,-2) {};
\node[] (q1) at (0.3,1) {$\bar q$};
\node[circle,draw=black, fill=white, inner sep=0pt,minimum size=5pt] (w1) at (3.5,1) {};
\draw (-1,-2) -- (5,-2);
\draw[fermion] (pi1) -- (R);
\draw[fermion] (pi1) -- (w1);
\draw[fermion] (R) -- (f);
\end{tikzpicture}
\caption{Example of a graph contributing to the trace formula for a cotangent target. Note there is no $\bar p$-derivative for the $R$-vertex. Moreover, one can check that it provides a correct degree count. Indeed, the amount of black vertices in the bulk is given by 2, hence $\dim \mathsf{C}^0_\Gamma(\mathbb{D})=4$ and the form degree of $\omega_\Gamma$ is given by $\vert R\vert+1+2=4$.}
\label{fig:ex1}
\end{figure}
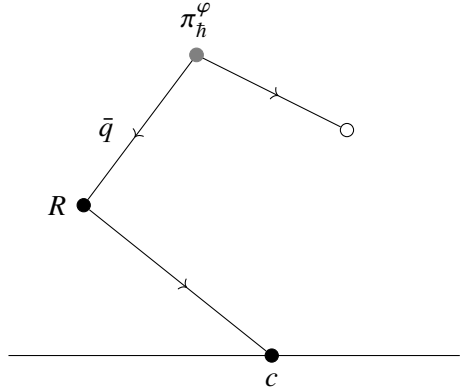

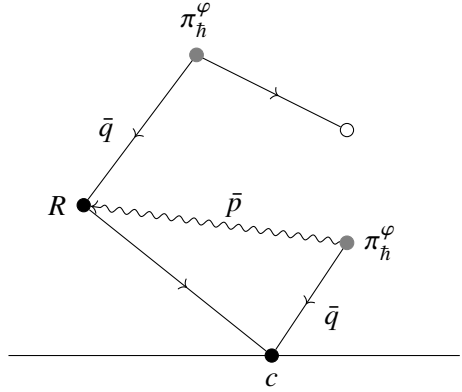
\begin{figure}[ht]
\centering
\begin{tikzpicture}
\tikzset{Bullet/.style={fill=black,draw,color=#1,circle,minimum size=0.5pt,scale=0.5}}
\node[vertex, label=left: $R$] (R) at (0,0) {};
\node[Bullet=gray, label=above: $\pi^\varphi_\hbar$] (pi1) at (1.5,2) {};
\node[Bullet=gray, label=right: $\pi^\varphi_\hbar$] (pi2) at (3.5, -0.5) {};
\node[vertex, label=below: $c$] (f) at (2.5,-2) {};
\node[] (q1) at (0.3,1) {$\bar q$};
\node[] (p) at (2,0) {$\bar p$};
\node[] (q2) at (3.3,-1.5) {$\bar q$};
\node[circle,draw=black, fill=white, inner sep=0pt,minimum size=5pt] (w1) at (3.5,1) {};
\draw (-1,-2) -- (5,-2);
\draw[fermion] (pi1) -- (R);
\draw[fermion] (pi1) -- (w1);
\draw[->,decorate,decoration={snake,amplitude=.4mm,segment length=2mm,post length=1mm}] (pi2) -- (R);
\draw[fermion] (pi2) -- (f);
\draw[fermion] (R) -- (f);
\end{tikzpicture}
\caption{Example of a graph contributing to the trace formula for a cotangent target. Note there is only one $\bar p$-derivative for the $R$-vertex. Moreover, one can check that it provides a correct degree count. Indeed, the amount of black vertices in the bulk is given by 3, hence $\dim \mathsf{C}^0_\Gamma(\mathbb{D})=6$ and the form degree of $\omega_\Gamma$ is given by $\vert R\vert+\vert\pi^\varphi_\hbar\vert+1+2=6$.}
\label{fig:ex2}
\end{figure}

\end{ex}

\subsection{Relation to the Nest--Tsygan theorem}
Let $M=T^*N$ be the cotangent bundle for a manifold $N$ endowed with its canonical symplectic form $\omega$ and consider the constant function $1$ on the boundary of the disk. In this setting we get the following theorem.

\begin{thm}
The trace formula \eqref{eq:globalized_trace} satisfies \eqref{eq:Nest-Tsygan}.
\end{thm}

\begin{figure}[ht]
\begin{tikzpicture}[scale=0.8]
\tikzset{Bullet/.style={fill=black,draw,color=#1,circle,minimum size=0.5pt,scale=0.5}}
\node[vertex] (1) at (0,0) {};
\node[vertex] (2) at (1,1) {};
\node[vertex] (3) at (3,1) {};
\node[vertex] (4) at (4,0) {};
\node[vertex] (5) at (4,-1.5) {};
\node[vertex] (6) at (3,-2.5) {};
\node[vertex] (7) at (1,-2.5) {};
\node[vertex] (8) at (0,-1.5) {};
\node[vertex] (center) at (2,-0.75) {};
\node[Bullet=gray] (pi1) at (-1,0.5) {};
\node[Bullet=gray] (pi2) at (0.5,1.8) {};
\node[Bullet=gray] (pi3) at (3.5,1.8) {};
\node[Bullet=gray] (pi4) at (5,0.5) {};
\node[Bullet=gray] (pi5) at (5,-2) {};
\node[Bullet=gray] (pi6) at (3.5,-3.5) {};
\node[Bullet=gray] (pi7) at (0.5,-3.5) {};
\node[Bullet=gray] (pi8) at (-1,-2) {};
\node[circle,draw=black, fill=white, inner sep=0pt,minimum size=5pt] (w1) at (-1,1.5) {};
\node[circle,draw=black, fill=white, inner sep=0pt,minimum size=5pt] (w2) at (1.3,2.5) {};
\node[circle,draw=black, fill=white, inner sep=0pt,minimum size=5pt] (w3) at (4.5,1.8) {};
\node[circle,draw=black, fill=white, inner sep=0pt,minimum size=5pt] (w4) at (5.5,-0.5) {};
\node[circle,draw=black, fill=white, inner sep=0pt,minimum size=5pt] (w5) at (5,-3) {};
\node[circle,draw=black, fill=white, inner sep=0pt,minimum size=5pt] (w6) at (2.5,-4) {};
\node[circle,draw=black, fill=white, inner sep=0pt,minimum size=5pt] (w7) at (-0.5,-3) {};
\node[circle,draw=black, fill=white, inner sep=0pt,minimum size=5pt] (w8) at (-1.5,-1) {};
\draw[fermion] (1) -- (2);
\draw[fermion] (2) -- (3);
\draw[fermion] (3) -- (4);
\draw[fermion] (4) -- (5);
\draw[fermion] (5) -- (6);
\draw[fermion] (6) -- (7);
\draw[fermion] (7) -- (8);
\draw[fermion] (8) -- (1);
\draw[fermion2] (pi1) -- (1);
\draw[fermion2] (pi2) -- (2);
\draw[fermion2] (pi3) -- (3);
\draw[fermion2] (pi4) -- (4);
\draw[fermion2] (pi5) -- (5);
\draw[fermion2] (pi6) -- (6);
\draw[fermion2] (pi7) -- (7);
\draw[fermion2] (pi8) -- (8);
\draw[fermion2] (pi1) -- (w1);
\draw[fermion2] (pi2) -- (w2);
\draw[fermion2] (pi3) -- (w3);
\draw[fermion2] (pi4) -- (w4);
\draw[fermion2] (pi5) -- (w5);
\draw[fermion2] (pi6) -- (w6);
\draw[fermion2] (pi7) -- (w7);
\draw[fermion2] (pi8) -- (w8);
\draw[fermion] (center) -- (1);
\draw[fermion] (center) -- (2);
\draw[fermion] (center) -- (3);
\draw[fermion] (center) -- (4);
\draw[fermion] (center) -- (5);
\draw[fermion] (center) -- (6);
\draw[fermion] (center) -- (7);
\draw[fermion] (center) -- (8);
\draw (2,-0.75) circle [radius=5cm];
\node[vertex,label=right: $1$] (b) at (7,-0.75) {};
\end{tikzpicture}
\caption{Example of a wheel graph that gives a contribution to the trace formula if we place the constant function $1$ on the boundary. The $\pi^\varphi_\hbar$-vertices are represented by the gray vertices and the $R$-vertices are represented by the black vertices. The picture without the center vertex and the corresponding arrows starting at the center is meant to be before merging. After merging we get the wheel with spokes pointing outwards.}
\label{fig:wheel_graph}
\end{figure}
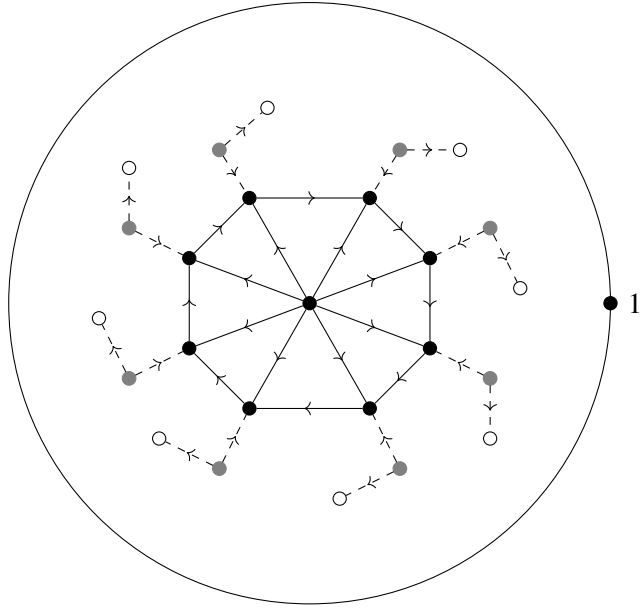

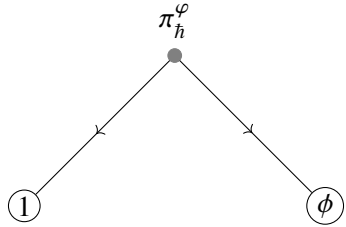
\begin{figure}[ht]
\centering
\begin{tikzpicture}
\tikzset{Bullet/.style={fill=black,draw,color=#1,circle,minimum size=0.5pt,scale=0.5}}
\node[Bullet=gray, label=above: $\pi^\varphi_\hbar$] (pi) at (0,0) {};
\node[residual] (res1) at (-2,-2) {$1$};
\node[residual] (res2) at (2,-2) {$\phi$};
\draw[fermion] (pi) -- (res1);
\draw[fermion] (pi) -- (res2);
\end{tikzpicture}
\caption{The appearing residual graphs. Here $1$ and $\phi$ both are regarded as the generators of the relative equivariant cohomology on the disk $H_{S^1}^\bullet(\mathbb{D},\de\mathbb{D})$.}
\label{fig:pi_graphs}
\end{figure}

\begin{proof}
One can easily check that by Proposition \ref{prop:linear_R} and degree reasons the only diagrams contributing within the trace formula are given by wheel-like loops as in Figure \ref{fig:wheel_graph}, and residual graphs as in Figure \ref{fig:pi_graphs}. Using the same construction as in Section \ref{subsec:tamrakin-tsygan}, we can merge the gray vertices to the center, and obtain wheel graphs which again will give rise to $\Hat{A}_u(TM)$. Recall that $\Hat{A}_u(TM)=\Hat{A}_0(TM)+u\Hat{A}_1(TM)+u^2\Hat{A}_2(TM)+\dotsm$, where $\Hat{A}_j(TM)\in H^{2j}(M)$. 
Note that we choose $\Omega$ to be the symplectic volume form $\frac{\omega^d}{d!}$ and, using \eqref{eq:Tamarkin-Tsygan}, we can see that if $c=1$, the $u$'s will all cancel eachother and thus it will not depend on $u$. Indeed, we have 
\begin{equation}
    \exp(\iota_\pi/u)=\sum_{n=0}^\infty\frac{1}{n!}\frac{1}{u^n}(\iota_\pi)^n=\sum_{n=0}^\infty\frac{\hbar^n}{n!}\frac{1}{u^n}\prod_{k=1}^n(\mathsf{T}\varphi^*\omega)^{i_kj_k}\left(\sum_{1\leq i_1<j_1<\dotsm <i_n<j_n\leq 2d}\prod_{k=1}^n\iota_{\de_{i_k}}\iota_{\de_{j_k}}\right),
\end{equation}
and therefore
\begin{equation}
    \exp(\iota_\pi/u)\frac{\omega^d}{d!}=\sum_{n= 0}^\infty\sum_{1\leq i<j\leq 2d} \frac{1}{u^n}\frac{\hbar^n}{n!d!}\prod_{k=1}^{d-n}(\mathsf{T}\varphi^*\omega)^{i_kj_k}\bigwedge_{k=1}^d\dd x^{i_k}\land \dd x^{j_k}.
\end{equation}
By degree reasons, the only surviving terms in $\Hat{A}_u(TM)\exp(\iota_\pi/u)\frac{\omega^d}{d!}$ are 
\begin{equation}
    \Hat{A}(TM)\exp(-\omega/\hbar)\prod_{k=1}^\infty\exp\left(\hbar^{k-1}\omega_k\right),\quad \omega_k\in \Omega^{2}(M)
\end{equation}

From the field theoretical construction, it is easy to check that the sum over all residual graphs will exactly give a contribution $\exp\left(\omega_\hbar/\hbar\right)$. Indeed, the integral 
\begin{equation}
\int_\mathbb{D}\phi(z,u)^{s}=\frac{\I}{2\pi}su^{s-1}\int_{\mathbb{D}}(1-\vert z\vert^2)^{s-1}\dd z\land \dd\bar z=u^{s-1},\quad s\geq 1,
\end{equation}
and for $s=1$, we get $\int_{\mathbb{D}}\phi=1$. Hence summing over all such graphs we get $\exp(\pi^\varphi_\hbar)=\exp(\omega_\hbar/\hbar)$. Putting everything together, we have 
\begin{equation}
    \tr(1)=\int_M\Hat{A}(TM)\exp(\omega_\hbar/\hbar).
\end{equation}

\end{proof}

\subsection{Reduction of the trace formula for cotangent targets}

\begin{prop}
The trace map for the globalized Poisson Sigma Model with cotangent target reduces to the trace map \eqref{eq:equiv_trace_map}.
\end{prop}

\begin{proof}
Consider the Poisson Sigma Model with target a cotangent bundle $M=T^*N$ for some manifold $N$ such that $\dim M=2d$. 
The Poisson structure is then induced by the canonical symplectic form $\omega$ on $M$. 
Note first that \eqref{eq:globalized_trace} can be written as 
\begin{equation}
    \label{eq:leading_order}
    \tr(f)=\int_M\rho(\mathsf{T}\varphi^*f)\vert_{y=0}\exp(\mathsf{T}\varphi^*h\vert_{y=0})\Omega+O(\hbar)=\int_Mf\exp(h)\Omega+O(\hbar), 
\end{equation}
where $h$ was the Hamiltonian function for $\pi$ such that $\Div_\Omega\pi-[h,\pi]=0$. Indeed, by considering the Feynman graph expansion of $\calV_n^{\pi^\varphi_\hbar}$, we get that 
\begin{equation}
    \tr(f)=\int_M\sum_{n=0}^\infty\frac{\hbar^n}{n!}P_n(\mathsf{T}\varphi^*\pi,\mathsf{T}\varphi^*h,\rho(\mathsf{T}\varphi^*f))\exp(\mathsf{T}\varphi^*h)\Omega,
\end{equation}
where $P_n$ are differential polynomials in $\mathsf{T}\varphi^*\pi,\mathsf{T}\varphi^*h$, and $\rho(\mathsf{T}\varphi^*f)$. 
Now, considering cotangent targets and choosing $\Omega$ to be the symplectic volume $\frac{\omega^d}{d!}$, we can see that the leading order $\hbar$ term of $\tr(f)$ is given by
\begin{equation}
   \int_Mf\frac{\omega^d}{d!}. 
\end{equation}
Since $\pi$ is constant, we get that $\Div_\Omega\pi=0$ and hence $[h,\pi]=0$, which implies that $h$ is constant, e.g. $h=0$. This is also compatible with the Nest--Tsygan theorem. One can compute 
\begin{multline}
    \tr(1)=\int_M\Hat{A}(TM)\exp(\omega_\hbar/\hbar)=\int_M(\Hat{A}_0(TM)+\Hat{A}_1(TM)+\dotsm)\sum_{n=0}^\infty \frac{1}{n!}\frac{\omega_\hbar^n}{\hbar^n}\\
    =\int_M(\Hat{A}_0(TM)+\Hat{A}_1(TM)+\dotsm)\sum_{n=0}^\infty\frac{1}{n!}\frac{1}{\hbar^n}\left(-\omega+\sum_{k\geq 1}\hbar^k\omega_k\right)^n\\
    =\frac{(-1)^d}{\hbar^d}\left(\int_M(\Hat{A}_0(TM)+\Hat{A}_1(TM)+\dotsm)\frac{\omega^d}{d!}+O(\hbar)\right)\\
    =\frac{(-1)^d}{\hbar^d}\left(\int_M\Hat{A}_0(TM)\frac{\omega^d}{d!}+O(\hbar)\right)=\frac{(-1)^d}{\hbar^d}\left(\int_M\frac{\omega^d}{d!}+O(\hbar)\right),
\end{multline}
Note that we have used $\Hat{A}_0(TM)=1\in H^0(M)$. 
Using the Feynman graphs for the corresponding effective theory together with the fact that, for a solution $\gamma$ of \eqref{eq:fedosov}, the leading $\hbar$ term of $\dd_{TM}\gamma$ is given by $\omega_{ij}\dd y^i\land \dd x^j$, it can be seen that the leading $\hbar$ term of \eqref{eq:equiv_trace_map} is given by 
\begin{equation}
    \int_M\int_{Ber}f\exp(\omega_{ij}\dd y^i\land\dd x^j/\hbar)=\frac{(-1)^d}{\hbar^d}\int_Mf\frac{\omega^d}{d!}.
\end{equation}
Here we use that the map $\int_{Ber}$ will give rise to the symplectic volume form $\frac{\omega^d}{d!}$ on $M$. Therefore the $\hbar$ leading terms coincide. Moreover, in \cite{GLL} they also show how the trace \eqref{eq:trace_symplectic} is compatible with the Nest--Tsygan theorem.
Note that the morphism $\calV^{\pi^\varphi_\hbar}$, which is given as the expectation value of the Fedosov-type formal global action, gives rise to the twisted integration map, where the effective action is indeed given in terms of a solution $\gamma$ of \eqref{eq:fedosov} since \eqref{eq:general_fedosov2} is reduced to \eqref{eq:fedosov} for the symplectic case as explained in Section \ref{subsec:Grothendieck_conn_symplectic_manifolds}. This action functional corresponds to $\gamma_\infty$, which can bee seen by using the corresponding Feynman rules on $S^1$.
Moreover, the Grothedieck connection gives rise to the globalization map $[\enspace]^{S^1}_\infty$ for observables and the quantization map $\rho$ reduces to the inverse of the symbol map $\sigma$. 
\end{proof}

\begin{appendix}

\section{BV algebras and relation to field theory}
\label{app_sec:BV_algebras}
We want to recall some notions on BV algebras as in \cite{Getzler1994}, and how it is related to the original gauge formalism developed by Batalin and Vilkovisky within quantum field theory.
\subsection{Braid algebras}
\label{app_subsec:braid_algebra}
Let us first recall what a braid algebra is. A braid algebra $B$ is a commutative DG algebra endowed with a Lie bracket $[\enspace,\enspace]$ of degree $+1$ satisfying the Poisson relations
\begin{equation}
    [a,bc]=[a,b]c+(-1)^{\vert a\vert(\vert b\vert -1)}b[a,c],\qquad \forall a,b,c\in B
\end{equation}
An identity element in $B$ is an element $\boldsymbol{1}$ of degree 0 such that it is an identity for the product and $[\boldsymbol{1},\enspace]=0$.
\subsection{BV algebras}
\label{app_subsec:BV_algebras}
A BV algebra $A$ is a commutative DG algebra endowed with an operator $\Delta\colon A_\bullet\longrightarrow A_{\bullet+1}$ such that $\Delta^2=0$ and 
\begin{align}
\label{app:eq:BV_Laplacian_2}
\begin{split}
    \Delta(abc)&=\Delta(ab)c+(-1)^{\vert a\vert}a\Delta(bc)+(-1)^{(\vert a\vert-1)\vert b\vert}b\Delta(ac)\\
    &-\Delta(a)bc-(-1)^{\vert a\vert}a\Delta(b)c-(-1)^{\vert a\vert+\vert b\vert}ab\Delta(c),\qquad \forall a,b,c\in A.
\end{split}
\end{align}
An identity in $A$ is an element $\boldsymbol{1}$ of degree 0 such that it is an identity for the product and $\Delta(\boldsymbol{1})=0$. One can show that a BV algebra is in fact a special type of a braid algebra. More precisely, a BV algebra is a braid algebra endowed with an operator $\Delta\colon A_\bullet\longrightarrow A_{\bullet+1}$ such that $\Delta^2=0$ and such that the bracket and $\Delta$ are related by
\begin{equation}
    [a,b]=(-1)^{\vert a\vert}\Delta(ab)-(-1)^{\vert a\vert}\Delta(a)b-a\Delta(b),\qquad\forall a,b\in A.
\end{equation}
Moreover, in a BV algebra we have 
\begin{equation}
    \Delta([a,b])=[\Delta(a),b]+(-1)^{\vert a\vert-1}[a,\Delta(b)],\qquad \forall a,b\in A.
\end{equation}
\subsection{Connection to field theory}
\label{app_subsec:connection_to_field_theory}
We would like to explain the name ``BV'' algebra. This comes from the approach to deal with gauge theories in quantum field theory developed by Batalin--Vilkovisky in the setting of odd symplectic (super)manifolds. Let $(\calF,\omega)$ be an odd symplectic (super)manifold. In physics, $\calF$ is called the space of fields.
Let $f\in C^\infty(\calF)$ and consider its Hamiltonian vector field $X_f$. One can check that $C^\infty(\calF)$ endowed with the Poisson bracket
\begin{equation}
\label{eq:odd_Poisson_bracket}
    \{f,g\}:=(-1)^{\vert f\vert -1}X_f(g)
\end{equation}
is a braid algebra. Let $\mu\in \Gamma(Ber(\calF))$ be a nowhere-vanishing section of the Berezinian bundle of $\calF$. This represents a density which is characterized by the integration map $\int\colon \Gamma_c(\calF,Ber(\calF))\longrightarrow \R$. Hence $\mu$ induces an integration map on functions with compact support
\begin{equation}
    C^\infty(\calF)\ni f\longmapsto \int_{\calL\subset\calF} f\mu^{1/2},
\end{equation}
for some Lagrangian submanifold $\calL\subset \calF$, where the integral exists.
Then one can define a divergence operator $\Div_\mu X$ by 
\begin{equation}
    \int_\calF (\Div_\mu X)f\mu=-\int_\calF X(f)\mu.
\end{equation}

\begin{lem}
For a vector field $X$ let $X^*=-X-\Div_\mu X$. Then 
\begin{equation}
    \int_\calF fX(g)\mu=(-1)^{\vert f\vert\vert X\vert}\int_\calF X^*(f)g\mu.
\end{equation}
Moreover, $\Div_\mu(fX)=f\Div_\mu X-(-1)^{\vert f\vert \vert X\vert}X(f)$ and if $S\in C^\infty(\calF)$ is an even function, then $\Div_{\exp(S)\mu}X=\Div_\mu X+X(S)$. 
\end{lem}
One can then define $\Delta$ to be the odd operator on $C^\infty(\calF)$ given by
\begin{equation}
\label{eq:Delta_div}
    \Delta(f)=\Div_\mu X_f.
\end{equation}
A BV (super)manifold $(\calF,\omega,\mu)$ is then an odd symplectic (super)manifold with Berezinian $\mu$ such that $\Delta^2=0$.
\begin{prop}
Let $(\calF,\omega,\mu)$ be a BV (super)manifold.
\begin{enumerate}
    \item The algebra $(C^\infty(\calF),\{\enspace,\enspace\},\Delta)$ is a BV algebra, where $\Delta$ is given as in \eqref{eq:Delta_div} and $\{\enspace,\enspace\}$ is the odd Poisson bracket coming from the odd symplectic form $\omega$ as in \eqref{eq:odd_Poisson_bracket}.
    \item The Hamiltonian vector field associated to some $f\in C^\infty(\calF)$ is given by the formula $X_f=-[\Delta,f]+\Delta(f)$, where $[\enspace,\enspace]$ denotes the commutator of operators.
    \item If $S\in C^\infty(\calF)$ and $\Delta_S$ is the operator associated to the Berezinian $\exp(S)\mu$, then $\Delta_S=\Delta-X_S$ and $\Delta_S^2=X_{\Delta(S)+\frac{1}{2}\{S,S\}}$.
\end{enumerate}
\end{prop}
Note that point (3) is exactly the case that we have in quantum field theory. Moreover, if 
\begin{equation}
\label{eq:QME_BV_alg}
    \Delta(S)+\frac{1}{2}\{S,S\}=0,
\end{equation}
we get that $\Delta_S^2=0$, which ensures a BV algebra structure. In physics, the function $S$ is called the action and Equation \eqref{eq:QME_BV_alg} is usually called the Quantum Master Equation\footnote{Here we have set $\I\hbar=1$, whereas in quantum field theory we want dependence on $\hbar$ as a formal variable and consider formal power series as Taylor expansions (cf. perturbative expansion of path integrals)}.

\end{appendix}

\printbibliography

\end{document}